\documentclass[10pt,twocolumn,journal]{IEEEtran}
\usepackage{latexsym}
\usepackage{amsfonts}
\usepackage{amsbsy}
\usepackage{amsmath,amssymb}
\usepackage{times}
\usepackage{graphicx}
\usepackage{enumerate}
\usepackage[usenames]{color}
\usepackage[dvips]{pstcol}
\usepackage{epstopdf}
\usepackage{cite}
\usepackage{amsmath}
\usepackage{amssymb}
\usepackage{amsfonts}
\usepackage{graphicx}
\usepackage{epsfig}
\usepackage{psfrag}
\usepackage{xcolor}
\usepackage{amsfonts, bm}
\usepackage{epstopdf}
\usepackage{cite}
\usepackage{color}
\usepackage{xcolor}
\usepackage{subfig}
\usepackage{stfloats}
\usepackage[linesnumbered,ruled]{algorithm2e}

\newtheorem{lemma}{\textbf{Lemma}}

\input epsf

\begin{document}
\title{Efficient Resource Allocation for Relay-Assisted Computation Offloading in Mobile Edge Computing}

\author{\IEEEauthorblockN{Xihan Chen, \IEEEmembership{Student Member,~IEEE}, Yunlong Cai, \IEEEmembership{Senior Member,~IEEE}, Qingjiang Shi, \IEEEmembership{Senior Member,~IEEE},\\
Min-Jian Zhao, \IEEEmembership{Member,~IEEE}, Benoit Champage, \IEEEmembership{Senior Member,~IEEE}, and Lajos Hanzo, \IEEEmembership{Fellow,~IEEE}}

\thanks{
X. Chen, Y. Cai and M. Zhao are with the Department of Information Science and Electronic Engineering, Zhejiang University, Hangzhou, China (e-mail: chenxihan@zju.edu.cn; ylcai@zju.edu.cn;  mjzhao@zju.edu.cn).

Q. Shi is with the School of Software Engineering, Tongji University, Shanghai, China (e-mail: qing.j.shi@gmail.com).

B. Champagne is with the Department of ECE, McGill University, Montreal,  Canada (e-mail: benoit.champagne@mcgill.ca).

L. Hanzo is with the Department of ECS, University of Southampton, Southampton, U.K. (e-mail:
lh@ecs.soton.ac.uk).
}
}

\maketitle

\begin{abstract}
In this article, we consider the problem of relay assisted computation offloading (RACO), in which user $A$ aims to share the results of computational tasks with another user $B$ through wireless exchange over a relay platform equipped with mobile edge computing capabilities, referred to as a mobile edge relay server (MERS). To support the computation offloading, we propose a hybrid relaying (HR) approach employing two orthogonal frequency bands, where the amplify-and-forward scheme is used in one band to exchange
computational results, while the decode-and-forward scheme is used in the other band to transfer the unprocessed tasks. The motivation behind the proposed HR scheme for RACO is to adapt the allocation of computing and communication resources both to dynamic user requirements and to diverse computational tasks. Within this framework, we seek to minimize the weighted sum of the execution delay and the energy consumption in the RACO system by jointly optimizing the computation offloading ratio,  the bandwidth allocation, the processor speeds, as well as the transmit power levels of both user $A$ and the MERS, under practical constraints on the available computing and communication resources.  The resultant problem is formulated as a non-differentiable and nonconvex optimization program with highly coupled constraints. By adopting a series of transformations and introducing auxiliary variables, we first convert this problem into a more tractable yet equivalent form. We then develop an efficient iterative algorithm for its solution based on the concave-convex
procedure. By exploiting the special structure of this problem, we also propose a simplified algorithm based on the inexact block coordinate descent method, with reduced computational complexity. Finally, we present numerical results that illustrate the advantages of the proposed algorithms over state-of-the-art benchmark schemes.
\end{abstract}

\IEEEpeerreviewmaketitle

\section{Introduction} \label{Intro}
Owing to the ever-increasing popularity of smart mobile devices, mobile data traffic continues to grow.  According to a recent study\cite{cisco}, the global mobile data traffic will grow at a compound annual growth rate of 46 percent from 2017 to 2022, reaching 77.5 exabytes per month by 2022. Meanwhile, the type of wireless services is also experiencing a major change, expanding from the traditional voice, e-mail and web browsing, to sophisticated applications such as augmented reality, face recognition and natural language processing, to name a few. These emerging services are both latency-sensitive and computation-intensive, hence requiring a reliable low-latency air interface and vast computational resources. In effect, the limited computational capability and battery life of mobile devices cannot guarantee the quality of user experience (QoE) expected for these new services.

To alleviate the performance bottleneck, mobile edge computing (MEC), a new network architecture that supports cloud computing along with Internet service  at the network edge, is currently the focus of great attention within the telecommunication industry. Due to the proxility of the mobile devices to the MEC server \cite{esti}, this architecture has the potential to significantly reduce latency, avoid congestion and prolong the battery lifetime of mobile devices by  running demanding applications and  processing tasks at the network edge, where ample computational and storage resources remain available \cite{1}. Recently, MEC has gained considerable interest within the research community \cite{2,3,4,5,stochastic1,stochastic2,power,6,game,dvfs,millimeterwave}. In \cite{3} and \cite{4}, the authors derived the optimal resource allocation solution for a single-user MEC system engaged in multiple elastic tasks, aiming to minimize the average execution latency of all tasks under power constraints.  A multi-user MEC system was considered in \cite{game}, where a game-theoretic model is employed to design computation offloading algorithms for both energy and latency minimization at mobiles. In \cite{2}, You {\em et al.} investigated the optimal resource and offloading decision policy for minimizing the weighted sum of mobile energy consumption under computation latency constraint in a multiuser MEC system based on either time division multiple access (TDMA) or orthogonal frequency division multiple access (OFDMA). Different from the deterministic task model considered in the above works, resource allocation strategies have also been developed for multi-user MEC systems under the stochastic task model, which is characterized by random task arrivals \cite{stochastic1,stochastic2,power}. For a multi-cell MEC system, the resource  management strategies for system performance improvement are more sophisticated. Sardellitti {\em et al.}~\cite{5} considered the joint optimization of radio and computational resources for computation offloading in a dense deployment scenario, in the presence of intercell interference. To overcome the performance bottleneck caused by the extremely high channel state information (CSI) signaling overhead in the centralized algorithm, Wang {\em et al.}~\cite{6} presented a decentralized algorithm based on the alternating direction method of multipliers (ADMM) for joint computation offloading, resource allocation and Internet content caching optimization in heterogeneous wireless networks with MEC. Research efforts have been devoted to the hardware design of MEC systems. For instance, Wang {\em et al.}~\cite{dvfs} investigated partial computation offloading using both dynamic voltage and frequency scaling\footnote{DVFS is a technique that varies the supply voltage and clock frequency of a processor based on the computation load, in order to provide the desired computation performance while reducing energy consumption. (DVFS) by considering either energy or latency minimization.}. Barbarossa {\em et al.}~\cite{millimeterwave} amalgamated MEC-based computation offloading techniques with millimeter wave (mmWave) communications and tackled the intermittency of mmWave links by relying on multiple links.

The aforementioned studies focus on a common scenario, where mobile terminals first offload their computational tasks to the MEC server, which then feeds back the results to the mobile terminals. In contrast to prior studies, we consider a \emph{relay-assisted computation offloading} (RACO) scenario, where user $A$ aims to share its computational results with another user $B$ through a relay platform equipped with MEC capabilities, referred to as a mobile edge relay server (MERS). An example of a chess gamming application under the proposed RACO architecture is shown in Fig. 1 to illustrate the practicality of the background scenario, where the entire process is divided into two phases, i.e., initialization and interaction. In the initialization phase, user $A$ and user $B$ first access the MERS, and the requested connections is confirmed by the game content server \cite{mobileGame1}. A game engine server is initialized by loading all users' account information and game data from the game content server, and then the game logic and user data are processed to render the raw game video \cite{mobileGame2}. Finally,  a game streaming server is activated to encode the generated raw game video, and the results are sent to each user via wireless links \cite{mobileGame3}.  In the interactive phase, user $A$ and user $B$ take some actions aiming to win the mobile game, where each action corresponds to a specific computational task. Moreover, the result of a specific action taken by any user is bound to significantly affect the final outcome of the mobile game. Hence, to actively participate in the game, each user take the corresponding action as a counterattack until knowing which action the opponent have taken.


For this type of scenarios, user $A$ can only perform a fraction of its tasks locally due to hardware limitations while the remaining fraction is transferred to the MERS, where more extensive resources are available. This example strongly motivates the need for an efficient and flexible relay schemes to support  computation offloading in MEC systems. Unfortunately, the existing relay schemes \cite{CWC} proposed for conventional wireless networks cannot be directly applied to MEC, due to the following reasons. First, different from the conventional wireless networks, the overall performance of MEC is substantially affected by both communication and computational aspects; hence, a novel design criterion that embraces both aspects should be considered. Second, there is a wide variety of emerging applications that could benefit from RACO, but for which the choice of relay scheme, i.e., amplify-and-forward (AF) versus decode-and-forward (DF), may have a major impact on performance and quality of experience. Motivated by the above considerations, we propose a novel hybrid relay (HR) architecture for the RACO system, to better support the exchange of computational results between different users. In this setup, a fraction of the available bandwidth is assigned to the AF relay scheme to transmit the locally computed results, while the remaining fraction is assigned to the DF relay scheme to transmit the offloaded raw data.

The proposed HR architecture offers several advantages over the existing AF and DF relay schemes, as it can inherit the benefits of both AF and DF schemes. In particular, the AF scheme tends to be superior at low signal-to-noise ratios (SNRs) due to its low  computational complexity and  reduced energy consumption and delay.
In contrast, the DF scheme tends to perform better at high SNRs, where it can mitigate the errors resulting from signal propagation between the source and the relay server \cite{CWC}.
  %
Hence, by combing the merits of AF and DF relay schemes to enhance the system performance, the HR architecture is suitable for a wider range of applications. Moreover, the additional design flexibility provided by the proposed HR architecture allows more efficient resource allocation, leading in turn to reduced execution delay and energy consumption (as demonstrated later). We emphasize that in our proposed approach, there is no need to explicitly carry out relay scheme selection as it will be automatically determined by the optimal offloading ratio.


\begin{figure}[!t]
\centering
\scalebox{0.32}{\includegraphics{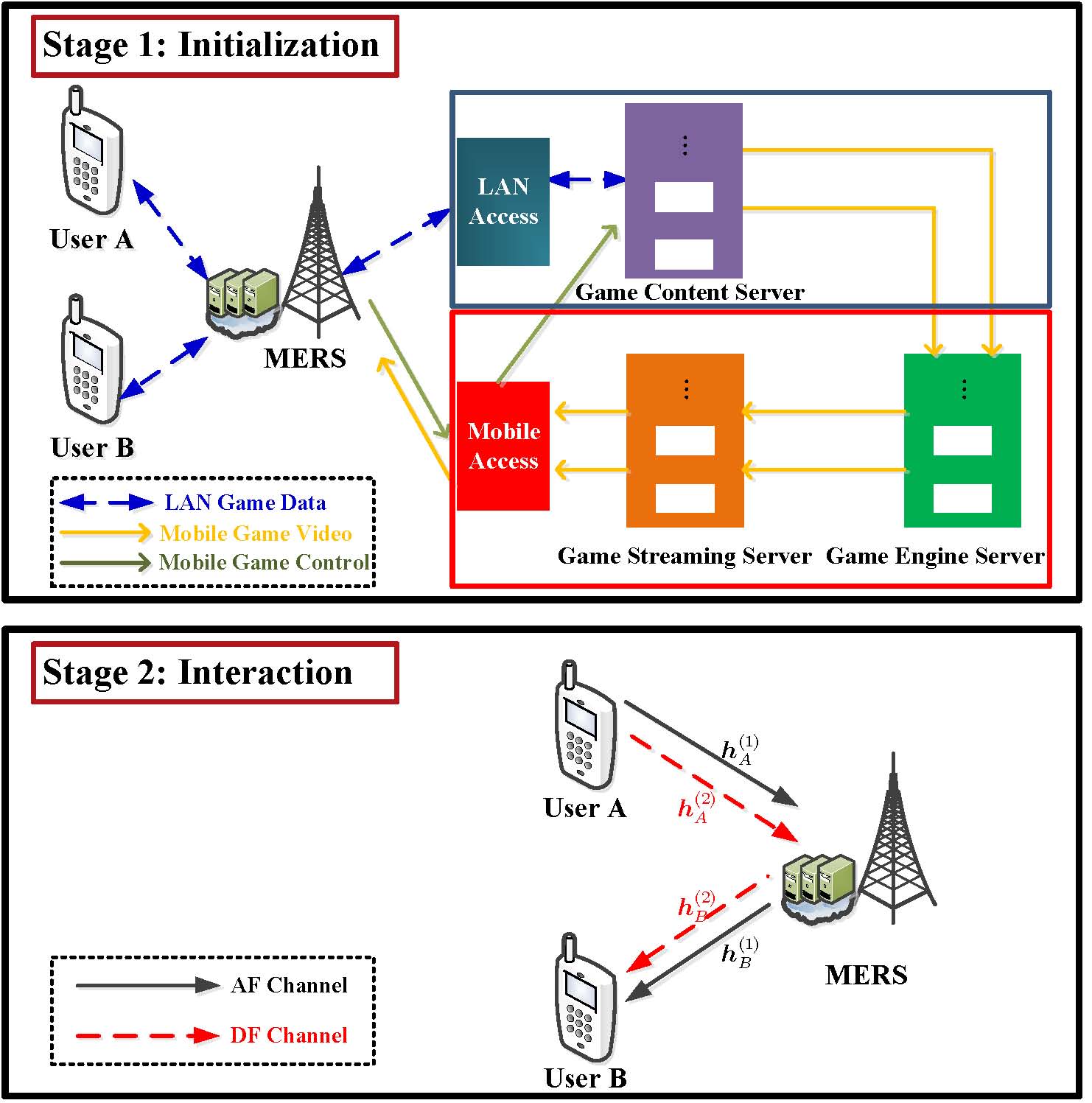}}
\caption{Illustration of a delay-sensitive mobile game application under the proposed RACO system with HR architecture.}\label{fig:game_illustraion}
\label{fig:graph}
\end{figure}

Within this framework, we seek to minimize the weighted sum of the execution delay and the energy consumption in the RACO system by jointly optimizing the computation offloading ratio, the processor clock rate, the bandwidth allocation between the AF and DF schemes, as well as the transmit power levels of user $A$ and the MERS, under the practical constraints imposed by the available computing and communication resources.
The formulated problem is very challenging due to the
highly coupled and non-differentiable nature of the  objective function and constraints. Still, by exploiting the structure of the problem and invoking the concave-convex procedure (CCCP) \cite{cccp}, we devise efficient joint resource allocation algorithms to solve it.

Against this background our main contributions can be summarized as follows:
\begin{itemize}
\item [1)] A new HR architecture is proposed for RACO system in MEC applications, i.e., to support computation offloading as well as the transfer of locally computed results. For this architecture, we formulate a joint resource allocation problem aiming to minimize the weighted sum of the execution delay and the energy consumption, subject to a number of realistic constraints.
\item [2)] By applying a series of suitable transformations and introducing auxiliary variables, we recast this challenging optimization problem into an equivalent but more tractable form. For the resultant problem, we develop a new CCCP-based algorithm to handle the highly coupled terms and to jointly optimize the RACO system parameters.
\item [3)] By further exploiting the problem structure, an efficient low-complexity algorithm is proposed based on the smooth approximation and the inexact block coordinate descent (IBCD) method of optimization.
\item [4)] To gain additional insights into the proposed RACO system with HR architecture, we consider two special cases for the off-loading ratio, respectively leading to AF only and DF only schemes. For these  cases, the optimization problem is further simplified and a pair of further algorithmic solutions are proposed.
\item [5)] Finally, we present and discuss our simulation results to shed more lights on both the convergence properties and the overall performance of our schemes and algorithms proposed for RACO in MEC applications.
\end{itemize}

This paper is structured as follows. Section \ref{sys_prb} describes the proposed RACO system model with HR architecture and formulates the resource allocation problem of interest. Section \ref{cccp_section} transforms the original problem into a more tractable yet equivalent form and develops a CCCP-based algorithm for its solution. In Section \ref{IBCD_section}, a smooth approximation is applied and a simplified algorithm based on the IBCD method is derived. Section \ref{Special case} considers the special cases of AF only and DF only, and for each case, proposes simplified and efficient solutions. Section \ref{simulation_section} presents the simulation results. Finally, the paper is concluded in Section \ref{conclusion_section}.

\section{System model and problem statement}\label{sys_prb}
We consider the RACO system with HR architecture illustrated in Fig. 1, where user $A$ aims to share the results of computational tasks with user $B$ through wireless exchange over a MERS, i.e., a relay platform equipped with MEC capabilities. Due to hardware limitations, user $A$ only performs a fraction of its tasks locally while the remaining fraction is transferred to the MERS, where more extensive resources are available. To support the computation offloading, we propose an HR architecture employing a combination of the AF and DF relay schemes over the entire available bandwidth.

Specifically, the AF scheme is used over a selected portion of the bandwidth to relay the computational results of user $A$ to user $B$, while the DF scheme is used over the remaining bandwidth to offload the computational tasks of user $A$ to the MERS and then to relay the results to user $B$. Clearly, depending on the task offloading ratio, the bandwidth allocation ratio, the available computing resources and the transmit power, user $A$'s offloading strategy may significantly affect both the end-to-end delay and the system's overall energy consumption. In this work, our aim is to balance these two system performance metrics by appropriately allocating both the computational and communication resources.

Denoting the total available spectral bandwidth by $W$ (in Hz), let $\nu\in[0,1]$ denote the
fraction of this bandwidth allocated to the DF scheme, so that $1-\nu$ is the fraction allocated to the AF scheme. Let $h_{A}^{(1)}$ and $h_B^{(1)}$ respectively denote the AF relay channels between user $A$ and the relay, as well as between the relay and user $B$. Similarly, let $h_{A}^{(2)}$ and $h_B^{(2)}$ respectively denote the DF relay channels between user $A$ and the relay, as well as between the relay and user $B$. Moreover, assume that partial offloading is implemented on the basis of the \emph{data partitioning-oriented tasks} of~\cite{110,dvfs}, and that the size of the computational results are proportional to the size of the input tasks. We characterize the computational tasks at user $A$  by the triplet ($L,K,\rho$), where $L$ (in bits) denotes the size of the tasks before computation, $K$ denotes the average number of central processing unit (CPU) cycles required for processing each bit, and  $\rho>0$ denotes the conversion ratio between the size of the tasks before computation and the size of the corresponding results after computation.
 Finally, let $\alpha \in [0, 1]$ denote the fraction of computational tasks offloaded by user A to the MERS, i.e.: $\alpha L$ bits are transferred to the MERS via the DF channel for remote processing, while $(1-\alpha)L$ bits are processed locally, and the results then being forwarded to the MERS via the AF channel.

First consider the AF relaying scheme. In this case, the received signal at the MERS is given by
\begin{equation}
y_R^{(1)}=\sqrt{P_{1}^{A}}h_{A}^{(1)} \overline{x}_{A1}+n_R^{(1)},
\end{equation}
where $\overline{x}_{A1}\sim \mathcal{CN}(0,1)$ denotes the transmit signal after local computing, $n_R^{(1)}\sim \mathcal{CN}(0,\sigma_{R1}^2)$ denotes the complex additive white Gaussian noise (AWGN) at the MERS, and $P_{1}^{A}$ denotes the transmit power of user $A$. The MERS  amplifies the received signal and forwards it to user $B$. Therefore, the signal received at user $B$ is given by
\begin{align} \label{1}
y_B^{(1)}{=}&\sqrt{P_{1}^{R}}h_{B}^{(1)}y_R^{(1)}+n_B^{(1)}\nonumber\\
=&\sqrt{P_{1}^{R}}\sqrt{P_{1}^{A}} h_{B}^{(1)}h_{A}^{(1)}\overline{x}_{A1}{+}\sqrt{P_{1}^{R}}h_{B}^{(1)}n_R^{(1)}+n_B^{(1)},
\end{align}
where $n_B^{(1)}\sim \mathcal{CN}(0,\sigma_{B1}^2)$ and $P_{1}^{R}$ denote the AWGN at user $B$ and the transmit power of the MERS allocated to the AF relaying scheme, respectively. According to \eqref{1}, the transmission rate and delay in the AF scheme are expressed as
\begin{align}
R_{\mathrm{AF}}&\triangleq\frac{(1-\nu)W}{2} \log_2\left(1+\frac { P_{1}^{A}P_{1}^{R}|h_{B}^{(1)} h_{A}^{(1)}|^2}{P_{1}^{R}|h_{B}^{(1)}|^2\sigma_{R1}^2+\sigma_{B1}^2}\right),\\
t_{\mathrm{AF}}&\triangleq\frac{(1-\alpha)\rho L}{R_{\mathrm{AF}}}.
\end{align}
Furthermore, the energy consumption of the AF scheme is given by
\begin{align}
E_{AF}\triangleq&(P_{1}^{A}+P_{1}^{R}\mathbb{E}[|y_R^{(1)}|^2])t_{\mathrm{AF}}\\
=&(P_{1}^{A}+P_{1}^{R}P_{1}^{A}|h_{A}^{(1)}|^2+P_{1}^{R}\sigma_{R1}^2)t_{\mathrm{AF}}.
\end{align}


Next, let us consider the DF relaying scheme. In this case, user $A$ first offloads its unprocessed computational tasks to the MERS, where they are decoded and processed. Similar to the AF relaying scheme, the signal received at the MERS is
\begin{align}
y_R^{(2)}&=\sqrt{P_{2}^{A}}h_{A}^{(2)} x_{A2}+n_R^{(2)},
\end{align}%
where $x_{A2}\sim \mathcal{CN}(0,1)$ denotes the transmit signal from user $A$, $n_R^{(2)}\sim \mathcal{CN}(0,\sigma_{R2}^2)$ denotes the AWGN at the MERS, and $P_{2}^{A}$ denotes the transmit power of user $A$ in the DF relay scheme. The transmission rate and delay for the first hop in the DF scheme are given by
\begin{align}
R_{\mathrm{DF1}}&\triangleq \nu W\log_2\left(1+\frac{ P_{2}^{A}| h_{A}^{(2)}|^2}{\sigma_{R2}^2}\right),\\
t_{\mathrm{DF1}}&\triangleq\frac{\alpha L}{R_{\mathrm{DF1}}}.
\end{align}
After decoding the message from $A$, the MERS executes the offloaded processing tasks and then re-encodes and forwards the computational results to user $B$. As in \cite{df2,df3}, it is assumed that the processing delay and energy consumption associated to the decoding and encoding operations at the MERS are negligible compared to those of edge computing, i.e. processing user $A$'s computational tasks. The signal received at user $B$ is given by
\begin{equation}\label{3}
y_B^{(2)}=\sqrt{P_{2}^{R}}h_{B}^{(2)} \overline{x}_{A2}+n_B^{(2)},
\end{equation}
where $\overline{x}_{A2}\sim \mathcal{CN}(0,1)$ denotes the transmit signal by the MERS after edge computing, $n_B^{(2)}\sim \mathcal{CN}(0,\sigma_{B1}^2)$ denotes the complex AWGN at the destination, and $P_{2}^{R}$ denotes the transmit power of MERS in the DF relaying scheme.  According to \eqref{3}, the rate and delay for the second hop in the DF scheme can be expressed as
\begin{align}
R_{\mathrm{DF2}}&\triangleq \nu W\log_2\left(1+\frac{ P_{2}^{R}| h_{B}^{(2)}|^2}{\sigma_{B2}^2}\right),\\
t_{\mathrm{DF2}}&\overline{}\triangleq\frac{\alpha \rho L}{R_{\mathrm{DF2}}}.
\end{align}
Furthermore, the corresponding energy consumption is
\begin{equation}
E_{DF}\triangleq P_{2}^{A}t_{\mathrm{DF1}}+P_{2}^{R}t_{\mathrm{DF2}}.
\end{equation}

As in \cite{dvfs}, we model the power consumption of a generic CPU as $P=\eta {F}^3$, where $F$ denotes the CPU's computation speed (in cycles per second) and $\eta$ is a coefficient depending on the chip architecture; hence, the energy consumption per cycle is given by $\eta {F}^2$. For local computation, the energy consumption can be minimized by optimally configuring computation speed via the DVFS technology \cite{vtc_EE}. Hence, if the amount of data bits processed at user $A$ is $(1-\alpha)L$, the execution time $t_l$ will be
$t_{l}\triangleq{K_l(1-\alpha)L}/{F_{l}},
$
where $F_l$ denotes the computation speed of user $A$, and the corresponding energy consumption $E_l$ is given by
$
E_l\triangleq(1-\alpha) L K_l \eta_l {F_l}^2.
$

Similarly, the execution time and the energy consumption of edge computation are respectively given by $t_{r}\triangleq{K_r \alpha L}/{F_{r}},$ and $E_{r}\triangleq \alpha L K_r \eta_r {F_r}^2,$
where $F_r$ denotes the computation speed of the MERS.

\setcounter{equation}{21}
\begin{figure*}[b]
\hrulefill
\begin{align}\label{lemma1_1}
\hat g(\bm{x}_1,\bm{x}_2;\bm{y}_1,\bm{y}_2)\!\triangleq\!\frac{1}{2}(h_1(\bm{x}_1)+h_2(\bm{x}_1))^2\!-\!\frac{1}{2}(h_1(\bm{y}_1)+h_2(\bm{y}_2))^2\!-\!
h_1^{'}(\bm{y}_1)h_1(\bm{y}_1)(\bm{x}_1-\bm{y}_1)-h_2^{'}(\bm{y}_2)h_2(\bm{y}_2)(\bm{x}_2-\bm{y}_2).
\end{align}
\end{figure*}

\setcounter{equation}{23}
\begin{figure*}[b]
\hrulefill
\begin{align}
f_1(\mathbf{x}, \bm{\phi})\triangleq& \frac{1}{2}[2L K_l \eta_l {F_l}^2\!+\! L K_r \eta_r (\alpha+{F_r}^2)^2\!+\!L K_l \eta_l (\alpha^2+{F_l}^4)\!+\!(P_{1}^{A}+t_{\mathrm{A}})^2\!+\!\sigma_{R1}^2(P_{1}^{R}+t_{\mathrm{A}})^2\nonumber\\
&+|h_{A}^{(1)}|^2(s_1+t_{\mathrm{A}})^2+(P_{2}^{A}+t_{\mathrm{D1}})^2+(P_{2}^{R}+t_{\mathrm{D2}})^2]+\gamma t_{s},\label{top1_1}\\
\hat f_2(\mathbf{x}, \bm{\phi};\tilde{\mathbf{x}},\tilde{\bm{\phi}})=&\frac{1}{2}[L K_r \eta_r(\tilde\alpha^2+\tilde{F_r}^4)+L K_l \eta_l(\tilde\alpha+\tilde{F_l}^2)^2+\!L K_r \eta_r \tilde\alpha(\!\alpha\!-\!\tilde\alpha)+2L K_r \eta_r \tilde{F_r}^3(\!F_r\!-\!\tilde{F_r})+L K_l \eta_l(\tilde\alpha+\tilde{F_l}^2)(\alpha-\tilde\alpha)\nonumber\\
&+2L K_l \eta_l\tilde{F_l}(\tilde\alpha+\tilde{F_l}^2)(F_l-\tilde{F_l})]+(\tilde{P_{1}^{A}})^2+{\tilde t_{\mathrm{A}}}^2+\sigma_{R1}^2((\tilde{P_{1}^{R}})^2+{\tilde t_{\mathrm{A}}}^2)+| h_{A}^{(1)}|^2(\tilde s_1^2+{\tilde t_{\mathrm{AF}}}^2)+({\tilde P_{2}^{A}})^2+{\tilde t_{\mathrm{D1}}}^2\nonumber\\
&+({\tilde P_{2}^{R}})^2+{\tilde t_{\mathrm{D2}}}^2]+[\tilde P_{1}^{A}(P_{1}^{A}-\tilde P_{1}^{A})+\tilde t_{\mathrm{A}}(t_{\mathrm{A}}-\tilde t_{\mathrm{A}})+\sigma_{R1}^2(\tilde P_{1}^{R}(P_{1}^{R}-\tilde P_{1}^{R})+\tilde t_{\mathrm{A}}(t_{\mathrm{A}}-\tilde t_{\mathrm{A}}))+\tilde P_{2}^{A}(P_{2}^{A}-\tilde P_{2}^{A})\nonumber\\
&+| h_{A}^{(1)}|^2(\tilde s_1(s_1-\tilde s_1)+\tilde t_{\mathrm{A}}(t_{\mathrm{A}}-\tilde t_{\mathrm{A}}))+\tilde t_{\mathrm{D1}}(t_{\mathrm{D1}}-\tilde t_{\mathrm{D1}})+\tilde P_{2}^{R}(P_{2}^{R}-\tilde P_{2}^{R})+\tilde t_{\mathrm{D2}}(t_{\mathrm{D2}}-\tilde t_{\mathrm{D2}})].\label{top1_2}
\end{align}
\end{figure*}

\setcounter{equation}{16}

Considering both AF and DF relaying schemes, the total latency for executing the computational tasks of user $A$ within the RACO framework is given by
\begin{equation}\label{t_system}
t_{\mathrm{sys}}\triangleq\mathrm{max} \{ t_{l}+t_{\mathrm{AF}},t_{\mathrm{DF1}}+t_{r}+t_{\mathrm{DF2}}\},
\end{equation}
 and the system's total energy consumption is expressed as
\begin{align}
E_{\mathrm{sys}}&\triangleq E_l+E_{r}+E_{AF}+E_{DF}.
\end{align}
Our interest in this work lies in finding efficient algorithmic solutions to the following resource allocation problem, referred to as the constrained \emph{weighted sum of execution delay and energy consumption minimization} problem\footnote{For constraints involving user index $i$, it is implicitly assumed that the constraint must apply $\forall i \in \{1,2\}$.}:
\begin{subequations}\label{WSDEM}
\begin{align}
    \mathbf{P1}: \mathop{\min}_{\mathbf{x}}&~ E_{\mathrm{sys}}+\gamma  t_{sys}\label{19a}\\
    \textrm{s.t.} \quad &0 < F_l\leq F^{\mathrm{max}}_{l},\label{19b}\\
    & 0 < F_r\leq F^{\mathrm{max}}_{r},\label{19c}\\
    & 0 \leq \alpha \leq 1,\label{19d}\\
        & 0 \leq \nu \leq 1,\label{19e}\\
       &  P_{i}^{A}\geq 0 , \label{19f}\\
      & P_{i}^{R}\geq 0, \label{19g}\\
      & \sum\limits_{i=1}^{2} P_{i}^{A} \leq P^{\mathrm{max}}_{A},\label{19h} \\
      & P_{1}^{R}\!\sigma_{R1}^2\!+\!| h_{A}^{(1)}\!|^2P_{1}^{R}P_{1}^{A}+P_{2}^{R} \leq P^{\mathrm{max}}_{R},\label{19i}
\end{align}
\end{subequations}
where $\mathbf{x}\triangleq [\alpha,\nu,{P}_{1}^{A},{P}_{2}^{A},{P}_{1}^{R},{P}_{2}^{R}, F_l,F_r]^T$ denotes the vector of search variables.
The objective function in $\mathbf{P1}$ is a weighted sum of the system's total latency and energy consumption, where the weighting factor $\gamma$ (in Joule/sec) allows a proper trade-off between these two key metrics.  Constraints \eqref{19b} and \eqref{19c} are the maximum computation speed constraints imposed by user $A$'s and the MERS's CPUs, respectively. Constraints \eqref{19e}, \eqref{19f}, \eqref{19g} and \eqref{19h} specify the transmission power budgets at user $A$ and the MERS.

Due to the nonconvex and non-differentiable objective function in \eqref{19a}, along with the nonconvex coupling constraint \eqref{19h}, the solution of problem $\mathbf{P1}$ remains challenging. In the following section, we propose a CCCP-based algorithm that can efficiently find a local stationary solution of problem $\mathbf{P1}$.

\section{Proposed CCCP Based Algorithm}\label{cccp_section}
In this section, we first transform problem $\mathbf{P1}$ into an equivalent yet more tractable form by introducing auxiliary variables, and subsequently develop an efficient CCCP based algorithm to solve the transformed problem. To this end, a locally tight upper bound is derived to obtain a convex approximation to the objective function, while linearization is applied to approximate the nonconvex constraints.
\setcounter{equation}{27}
\begin{figure*}[b]
\hrulefill
\begin{subequations}\label{ccccp}
\begin{align}
    \mathop{\min}_{\mathbf x, \bm{\phi}}&~ f_3(\mathbf x,\bm{\phi};\tilde{\mathbf x},\tilde{\bm{\phi}})\\
    \textrm{s.t.} ~
    &\Lambda_{\mathrm{A}}-\log_2(1+{1}/{\tilde\varphi_1})+(\varphi_1-\tilde \varphi_1)/[{\ln2}({(\tilde\varphi_1)^2+\tilde\varphi_1})]\leq0,\\
    &{2}R_{\mathrm{A}}+{W}(\nu+\Lambda_{\mathrm{A}})^2/{2}-W\Lambda_{\mathrm{A}}
-{W}((\tilde \nu)^2+(\tilde \Lambda_{\mathrm{A}})^2)/{2}-W{\tilde \nu}(\nu-\tilde \nu)-W{\tilde \Lambda_{\mathrm{A}}}(\Lambda_{\mathrm{A}}-\Lambda_{\mathrm{A}})\leq0,\\
    &2\alpha L\!+\!t_{\mathrm{D1}}^2\!+\!R_{\mathrm{D1}}^2\!-\!(\tilde R_{\mathrm{D1}}\!+\!\tilde t_{\mathrm{D1}})^2\!-\!2(\tilde R_{\mathrm{D1}}\!+\!\tilde t_{\mathrm{D1}})(R_{\mathrm{D1}}\!+\!t_{\mathrm{D1}}\!-\!\tilde R_{\mathrm{D1}}\!-\!\tilde t_{\mathrm{D1}})\leq 0,\\
    &R_{\mathrm{D1}}\!+\!{W}(\nu^2\!+\!\Lambda^2_{\mathrm{D1}})\!/{2}\!-\!{W}(\tilde\nu+\tilde \Lambda_{\mathrm{D1}})^2\!/{2}\!-\!W(\tilde\nu\!+\!\tilde \Lambda_{\mathrm{D1}})(\nu\!+\!\Lambda_{\mathrm{D1}}\!-\!\tilde\nu\!-\!\tilde \Lambda_{\mathrm{D1}})\!\leq0,\\
    &\Lambda_{\mathrm{D1}}-\log_2(1+{1}/{\tilde\varphi_2})+(\varphi_2-\tilde \varphi_2)/[{\ln2}({(\tilde\varphi_2)^2+\tilde\varphi_2})]\leq0,\\
    &2\sigma_{R2}^2+| h_{A}^{(2)}|^2[\varphi_2^2+{P_{2}^{A}}^2-(\tilde\varphi_2+\tilde P_{2}^{A})^2-2(\tilde\varphi_2+\tilde P_{2}^{A})(\varphi_2+P_{2}^{A}-\tilde\varphi_2-\tilde P_{2}^{A})]\leq 0,\\
    &  2\alpha \rho L\!+\!t_{\mathrm{D2}}^2\!+\!R_{\mathrm{D2}}^2\!-\!(\tilde R_{\mathrm{D2}}\!+\!\tilde t_{\mathrm{D2}})^2\!\!-\!\!2(\tilde R_{\mathrm{D2}}\!+\!\tilde t_{\mathrm{D2}})(R_{\mathrm{D2}}\!+\!t_{\mathrm{D2}}\!-\!\tilde R_{\mathrm{D2}}\!-\!\tilde t_{\mathrm{D2}})\!\leq \!0,\\
    &R_{\mathrm{D1}}\!+\!{W}(\nu^2\!+\!\Lambda^2_{\mathrm{D2}})\!/{2}\!-\!{W}(\tilde\nu+\tilde \Lambda_{\mathrm{D2}})^2\!/{2}\!-\!W(\tilde\nu\!+\!\tilde \Lambda_{\mathrm{D2}})(\nu\!+\!\Lambda_{\mathrm{D2}}\!-\!\tilde\nu\!-\!\tilde \Lambda_{\mathrm{D2}})\!\leq0,\\
    &\Lambda_{\mathrm{D2}}-\log_2(1+{1}/{\tilde\varphi_3})+(\varphi_3-\tilde \varphi_3)/[{\ln2}({(\tilde\varphi_3)^2+\tilde\varphi_3})]\leq0,\\
    &2\sigma_{B2}^2+| h_{B}^{(2)}|^2[\varphi_3^2+{P_{2}^{R}}^2-(\tilde\varphi_3+\tilde P_{2}^{R})^2-2(\tilde\varphi_3+\tilde P_{2}^{R})(\varphi_3+{P_{2}^{R}}-\tilde\varphi_3-\tilde P_{2}^{R})]\leq 0,\\
    &(P_{1}^{R}+P_{1}^{A})^2-(\tilde{P_{1}^{R}})^2-(\tilde{P_{1}^{A}})^2-2{\tilde{P_{1}^{R}}}({P_{1}^{R}}-\tilde{{P_{1}^{R}}})-2{\tilde{P_{1}^{A}}}({P_{1}^{A}}-\tilde{{P_{1}^{A}}})-2s_1\leq 0,\\
    &2s_2+{P_{1}^{R}}^2+{P_{1}^{A}}^2- (\tilde P_{1}^{A}+\tilde P_{1}^{R})^2-2(\tilde P_{1}^{A}+\tilde P_{1}^{R})(P_{1}^{A}+P_{1}^{R}-\tilde P_{1}^{A}-\tilde P_{1}^{R})\leq 0,\\
    &2K_l(1-\alpha)L+F_{l}^2+t_{l}^2-(\tilde{t_{l}}+\tilde{F_{l}})^2-2(\tilde{t_{l}}+\tilde{F_{l}})(t_{l}+F_{l}-\tilde{t_{l}}-\tilde{F_{l}})\leq 0,\\
    &2K_r\alpha L+F_{r}^2+t_{r}^2-(\tilde{t_{r}}+\tilde{F_{r}})^2-2(\tilde{t_{r}}+\tilde{F_{r}})(t_{r}+F_{r}-\tilde{t_{r}}-\tilde{F_{r}})\leq 0,\\
    &\eqref{19b}-\eqref{19h},\eqref{20h},\eqref{ccccp1}.
\end{align}
\end{subequations}
\end{figure*}

\setcounter{equation}{19}
\subsection{Problem Transformation}\label{cccp_suba}
We first transform problem $\mathbf{P1}$ into an equivalent but more tractable form. Specifically, by introducing a number of auxiliary variables represented by the vector $\bm{\phi}\triangleq [ t_{s},t_{\mathrm{A}},t_{\mathrm{D1}},t_{\mathrm{D2}},t_{l},t_{r},R_{\mathrm{A}},R_{\mathrm{D1}},R_{\mathrm{D2}},\Lambda_{\mathrm{A}},\Lambda_{\mathrm{D2   }},\Lambda_{\mathrm{D2}},\varphi_1,\varphi_2,\\$$\varphi_3,s_1,s_2]^T$, problem $\mathbf{P1}$ can be formulated as the following equivalent problem:
\begin{subequations}
\begin{align}
    \mathbf{P2}: \mathop{\min}_{\mathbf{x}, \bm{\phi}}&~  E_{\mathrm{sys}}+\gamma t_{s}\label{119a}\\
    \textrm{s.t.} \quad
    &
    (1-\alpha)\rho L\leq t_{\mathrm{A}}R_{\mathrm{A}},
    ~2 R_{\mathrm{A}}\leq (1-\nu)W \Lambda_{\mathrm{A}} ,\label{bb1}\\
    &\Lambda_{\mathrm{A}} \leq \log_2(1+ {1}/{\varphi_1}),
    ~\alpha L\leq t_{\mathrm{D1}}R_{\mathrm{D1}},\\
    & R_{\mathrm{D1}}\leq \nu W \Lambda_{\mathrm{D1}},
    ~\Lambda_{\mathrm{D1}} \leq \log_2(1+{1}/{\varphi_2}),\label{bb2}\\
    & \alpha \rho L\leq{R_{\mathrm{D2}}}t_{\mathrm{D2}},
    ~R_{\mathrm{D2}}\leq \nu W \Lambda_{\mathrm{D2}},\\
    &\Lambda_{\mathrm{DF2}} \leq \log_2(1+{1}/{\varphi_3}),
   \sigma_{R2}^2-\varphi_2 P_{2}^{A} |h_{A}^{(2)}|^2\leq 0,\\
   & |h_{B}^{(1)}|^2\sigma_{R1}^2P_{1}^{R}+\sigma_{B1}^2\!-\!|h_{B}^{(1)} h_{A}^{(1)}|^2 \varphi_1 s_2\leq 0, \\
    & \sigma_{B2}^2-\varphi_3 P_{2}^{R} |h_{B}^{(2)}|^2\leq 0, s_2 \leq P_{1}^{A}P_{1}^{R}\leq s_1,\\
    & K_l(1-\alpha)L\leq t_{l}{F_{l}},~K_r\alpha L \leq t_{r}{F_{r}}, \\
     &t_{l}+t_{\mathrm{A}} \leq t_{s},~t_{\mathrm{D1}}+t_{r}+t_{\mathrm{D2}}\leq t_{s},\label{20h}\\
     &\eqref{19b}-\eqref{19h},
\end{align}
\end{subequations}
where
\begin{align}
E_{\mathrm{sys}}\triangleq& (1\!-\!\alpha) L K_l \eta_l {F_l}^2+\!\alpha L K_r \eta_r {F_r}^2 t_{\mathrm{A}}+\!P_{2}^{A}t_{\mathrm{D1}}+P_{2}^{R}t_{\mathrm{D2}}\nonumber\\
&+(P_{1}^{A}+P_{1}^{R}\sigma_{R1}^2+s_1| h_{A}^{(2)}|^2)
\end{align}
denotes the system energy consumption. Note that problem $\mathbf{P1}$ and $\mathbf{P2}$ share the same global optimal solution for $\mathbf x$ under the given constraints. The detailed derivation of the equivalence between problems $\mathbf{P1}$ and $\mathbf{P2}$ is presented in Appendix \ref{appendix_A}.

\subsection{Proposed Algorithm for Solving Problem $\mathbf{P2}$}\label{cccp_subb}
In this part, we propose an efficient CCCP-based algorithm for solving problem $\mathbf{P2}$. In order to approximate this problem as a convex one, we first find a locally tight upper bound of the objective and then linearize the nonconvex constraints with the aid of the CCCP concept, so that the nonconvex problem $\mathbf{P2}$ can be approximated as a convex one.

\subsubsection{Upper Bound for the Objective Function}
Our approach for bounding the objective function in problem P2 relies on the following lemma.
\begin{lemma}{\label{lemma1}}
\cite{parallel} Supposes that $g$ has a separable structure as the product of two convex and non-negative real-valued
functions $h_1$ and $h_2$, that is,
$
    g(\bm{x}_1,\bm{x}_2)=h_1(\bm{x}_1)h_2(\bm{x}_2).
$
For any $(\bm{y_1},\bm{y_2})$ in the domain of $g$, a convex approximation of $g(\bm{x}_1,\bm{x}_2)$ in the neighborhood of $(\bm{y}_1,\bm{y}_2)$, which satisfies mild conditions required by the CCCP algorithm, is defined in \eqref{lemma1_1} as displayed at the bottom of this page.
\end{lemma}

Note that the continuity and smoothness conditions of the CCCP requires the strongly convex approximation of the objective function to have the same first derivative as the objective function, while the convex approximations of the constraints are required to be tight at the point of interest and to  bound the original constraints. Based on \emph{Lemma 3.1}, we can obtain a locally tight upper bound for the objective function of problem $\mathbf{P2}$ in the current iteration as follows,
\setcounter{equation}{22}
\begin{align}\label{f30}
    f_3(\mathbf{x}, \bm{\phi};\tilde{\mathbf{x}},\tilde{\bm{\phi}})\triangleq f_1(\mathbf x,\bm{\phi})-\hat f_2(\mathbf{x}, \bm{\phi};\tilde{\mathbf{x}},\tilde{\bm{\phi}}    ).
\end{align}
where $\tilde{\bm{x}}\triangleq [\tilde{\alpha},\nu,\tilde{{P}}_{1}^{A},\tilde{{P}}_{2}^{A},\tilde{{P}}_{1}^{R},\tilde{{P}}_{2}^{R}, \tilde{F}_l,\tilde{F}_r]^T$, and $\tilde{\bm{\phi}}\triangleq [ \tilde{t}_{s},\tilde{t}_{\mathrm{A}},\tilde{t}_{\mathrm{D1}}, \tilde{t}_{\mathrm{D2}},\tilde{t}_{l},\tilde{t}_{r}, \tilde{R}_{\mathrm{A}},\tilde{R}_{\mathrm{D1}},\tilde{R}_{\mathrm{D2}},\tilde{\varphi}_1,\tilde{\varphi}_2,\tilde{\varphi}_3,\tilde{s}_1,\tilde{s}_2]^T$ are the current points generated by the last iteration; $f_1(\mathbf{x}, \bm{\phi})$ and $\hat f_2(\mathbf{x}, \bm{\phi};\tilde{\mathbf{x}},\tilde{\bm{\phi}})$ are respectively defined in \eqref{top1_1}-\eqref{top1_2} as displayed at the bottom of this page.

Please refer to Appendix \ref{appendix_B} for the constructive derivation.

\subsubsection{Linearizing the Nonconvex Constraints}
Note that all the nonconvex constraints in problem $\mathbf{P2}$ have a similar structure and can be equivalently converted to convex constraints. Here, we focus on the conversion of  the first constraint in \eqref{bb1} as an example. By applying the equality $xy=\frac{1}{2}[\left(x+y\right)^2-x^2-y^2]$, we can rewrite this constraint into the DC program:
\setcounter{equation}{25}
\begin{align}
\underbrace{2(1-\alpha)\rho L+t_{\mathrm{AF}}^2+R_{\mathrm{AF}}^2}_{\mathrm{convex \quad function}}-\underbrace{(R_{\mathrm{AF}}+t_{\mathrm{AF}})^2}_{\mathrm{convex \quad function}}\leq 0,\label{ccc1}
\end{align}
By linearizing the subtracted convex terms in \eqref{ccc1} by applying the first-order Taylor expansion around the current point $(\tilde{\mathbf{x}},\tilde{\bm{\phi}})$, we obtain
\begin{align}\label{ccccp1}
&2(1-\alpha)\rho L+t_{\mathrm{A}}^2+R_{\mathrm{A}}^2-(\tilde R_{\mathrm{A}}+\tilde t_{\mathrm{A}})^2\nonumber\\
&\quad\quad\quad\quad-2(\tilde R_{\mathrm{A}}+\tilde t_{\mathrm{A}})(R_{\mathrm{A}}+t_{\mathrm{A}}-\tilde R_{\mathrm{A}}-\tilde t_{\mathrm{A}})\leq 0.
\end{align}
The other constraints in problem $\mathbf{P2}$ can be converted by a similar process, but the details are omitted due to space limitation. Finally, based on the CCCP concept, problem $\mathbf{P2}$ can be reformulated as an iterative sequence of  convex optimization problems defined in \eqref{ccccp} as displayed at the
bottom of this page.

Problem \eqref{ccccp} can be efficiently solved by the convex programming toolbox CVX \cite{cvx}. The implementation of Algorithm 1  is
summarized as Algorithm 1.  Repeated application of the CCCP iteration will eventually lead to a stationary solution of problem $\mathbf{P2}$ \cite{convergenceofCCCP}. We can show that the limit point of the iterates generated by Algorithm 1 also satisfies the KKT conditions of the DC program \eqref{ccccp}, which guarantees convergence to a local optimal solution of problem $\mathbf{P1}$. The proof is similar to that of \emph{Lemma 2} and \emph{Theorem 1} in \cite{cccp_convergence}, and we thus omit the details. The overall computational complexity of Algorithm 1 is dominated by the interior point method implemented by CVX toolbox, which is significantly affected by the number of second-order cone (SOC) constraints of problem \eqref{ccccp} and the corresponding dimensions. To access the complexity, we transform the constraints of problem \eqref{ccccp} into the form of SOC (details are omitted due to space limitation). In a nutshell, problem \eqref{ccccp} contains 7 SOC constraints of dimension 3 while the number of optimized variables is 21. Hence, the number of required floating point operations (FPOs) at each iteration is on the order of $10^5$.  The complexity of Algorithm 1 is given by the number of  required FPOs $n_1=M_1 I_1$, where $I_1$ denotes the number of required iterations and $M_1$ denotes the number of  FPOs at each iteration.
\begin{algorithm}[h]
\scriptsize
\centering
\caption{The CCCP-based iterative algorithm}\label{table1}
\begin{itemize}
\item [0.] \textbf{Initialization}: Define the tolerance of accuracy $\delta$ and the maximum number of iterations $N_{\text{max}}$. Initialize the algorithm with a feasible point $\mathbf{x}^{0},\bm{\phi}^{0}$. Set the iteration number $i=0$.
\item [1.]\textbf{repeat}
\item [2.]\quad Solve the convex optimization problem \eqref{ccccp} with the affine approximation, and assign the solution to $\mathbf{x}^{i+1},\bm{\phi}^{i+1}$.
\item [3.]\quad  Update the iteration number: $i\leftarrow i+1$
\item [4.]\textbf{until} $|  f_3(\mathbf{x}^{i},{\bm{\phi}}^{i})- f_3(\mathbf{x}^{i-1},{\bm{\phi}}^{i-1})|\leq \delta$ or reaching the maximum iteration number.
\end{itemize}
\end{algorithm}
\section{Proposed Low complexity Algorithm for Problem $\mathbf{P1}$}\label{IBCD_section}
The proposed CCCP-based algorithm  can be applied to address problem $\mathbf{P1}$ but incurs a very high computational complexity due to the need to solve the sequence of convex optimization problems \eqref{ccccp}. In this section, by further exploiting the problem structure, we propose an alternative iterative algorithm with much reduced complexity. Specifically, we first approximate the objective function of P1 as a smooth function and then propose an inexact  BCD algorithm (a variant of the BCD algorithm \cite{IBCD2017}) to solve the resulting problem.
\subsection{Smooth Approximation of Objective Function}
Observing that the constraints in $\mathbf{P1}$ are separable w.r.t. the five blocks of variables, i.e., $F_l$, $F_r$, $\alpha$, $\nu$, and $\mathbf{y}\triangleq\{{P}_{1}^{A},{P}_{2}^{A},{P}_{1}^{R},{P}_{2}^{R}\}$, we may apply the IBCD algorithm to problem $\mathbf{P1}$. This requires the objective function to be differentiable, which is not the case here due to \eqref{t_system}. To address the nondifferentiability issue, we first approximate the objective function of $\mathbf{P1}$ as a smooth function using the log-smooth method. Specifically, using the log-sum-exp inequality \cite[pp. 72]{cvx_book}, we have
\begin{equation}
\max(x,y){\leq}\frac{1}{\beta}\log(\exp(\beta x)+\exp(\beta y))\leq \max(x,y)+\frac{1}{\beta}\log 2. \label{log_inequality}
\end{equation}
Utilizing \eqref{log_inequality}, we can approximate $t_{\mathrm{sys}}$ as
$$\hat t_{\mathrm{sys}}\approx \frac{1}{\beta}\log(\exp(\beta (t_{l}+t_{\mathrm{AF}}))+\exp(\beta (t_{\mathrm{DF1}}+t_{r}+t_{\mathrm{DF2}})))$$ with a large $\beta$. Problem $\mathbf{P1}$ is smoothly approximated as
\begin{align}\label{smoothProblem}
    \mathop{\min}_{\mathbf{x}}~ f_{\beta}(\mathbf{x})
    \quad\textrm{s.t.} \quad \eqref{19b}-\eqref{19h},
\end{align}
where $f_{\beta}(\mathbf{x})\triangleq E_{\mathrm{sys}}+\gamma \hat t_{sys}$ denotes the approximated objective function of problem $\mathbf{P1}$ with the differentiable property.
\subsection{Inexact Block Coordinate Algorithm for Smoothed Problem}

We can now use the IBCD method to solve the smoothed problem \eqref{smoothProblem}. In this method, we sequentially update each block of variables, while fixing the other blocks to their previous values. For problem \eqref{smoothProblem}, this amounts to the following steps:

\emph{Step 1: Updating $F_l$ while fixing $\{F_r,\alpha, \nu, \mathbf{y}\}.$} Let us consider the subproblem w.r.t. $F_l$, which is given by
\begin{align}\label{subp_fl}
\min_{0 < F_l\leq F^{\mathrm{max}}_{l}}& \, E_l+\gamma \hat t_{sys}
\end{align}
It can be verified that the above subproblem is convex, and thus can be easily solved using the bisection method\cite{cvx_book}.

\emph{Step 2: Updating $F_r$ while fixing $\{F_l,\alpha,\nu,\mathbf{y}\}.$} Similar to the subproblem w.r.t. $F_l$, the subproblem w.r.t $F_r$ is also convex and thus can be solved using the bisection method.

\emph{Step 3: Updating $\alpha$ while fixing $\{F_l,F_r,\nu,\mathbf{y}\}.$} The subproblem w.r.t. $\alpha$ is also convex and thus can be solved using bisection.

\emph{Step 4: Updating $\nu$ while fixing $\{F_l,F_r,\alpha,\mathbf{y}\}.$} Since the subproblem w.r.t. $\nu$ is the minimization of a scalar function, it can be efficiently solved by the line-search method\cite{cvx_book}.

\emph{Step 5: Updating $\mathbf{y}$ while fixing $\{F_l, F_r,\nu,\alpha\}.$} Let us consider the subproblem w.r.t. $\mathbf{y}$, which is given by
\begin{align}\label{p1}
&\min_{\mathbf{y}} f_{\beta}(\mathbf{y})\quad
 \textrm{s.t.} \quad \eqref{19e}-\eqref{19h}.
\end{align}
Obviously, \eqref{19h} is a nonconvex constraint, which complicates the solution of \eqref{p1}. To efficiently update $\mathbf{y}$ while decreasing the objective value, we apply the concept of linearization to tackle the nonconvexity of \eqref{19h}. First, we express the latter as a DC program:
\begin{align}
P_{2}^{R}{+}P_{1}^{R}\sigma_{R1}^2\!{+}\frac{1}{2}| h_{A}^{(1)}|^2[(P_{1}^{R}\!{+}P_{1}^{A})^2{-}({P_{1}^{R}})^2{-}({P_{1}^{A}})^2] \leq P_R^{\mathrm{max}}.\!\label{dc}
\end{align}
By linearizing the nonconvex term $-({P_{1}^{R}})^2-({P_{1}^{A}})^2$ at the current point $\mathbf{\tilde y} \triangleq[\tilde{P}_{1}^{A},\tilde{P}_{2}^{A},\tilde{P}_{1}^{R}, \tilde{P}_{2}^{R}]^T$, we approximate \eqref{dc} as a convex constraint
\begin{align}\label{22}
U(\mathbf{y};\mathbf{\tilde y})\!\triangleq& \!P_{2}^{R}\!+\!P_{1}^{R}\sigma_{R1}^2\!+\!\frac{1}{2}| h_{A}^{(1)}|^2[(P_{1}^{R}+P_{1}^{A})^2\!-\!(\tilde{P_{1}^{R}})^2\nonumber\\
&-\!(\tilde{P_{1}^{A}})^2\!-2{\tilde{P_{1}^{R}}}\!{P_{1}^{R}}\!\!-\!2{\tilde{P_{1}^{A}}}{P_{1}^{A}}]\!-\!P_R^{\mathrm{max}}\leq 0.
\end{align}
As a result, we can approximate problem \eqref{p1} as
\begin{align}\label{p2}
\min_{\mathbf{y}} f_{\beta}(\mathbf{y})\quad
\textrm{s.t.}~~\eqref{19e}-\eqref{19g}, \eqref{22},
\end{align}
where the constraints are now all convex. Hence, we can apply the \emph{one-step} projected gradient (PG) method\cite{cvx_book} to problem \eqref{p2}.  Specifically, we update $\mathbf{y}$ according to 
\begin{align}
    \overline {\mathbf{y}}&=P_\Omega[\mathbf{\tilde y}-\nabla f_{\beta}(\mathbf{\tilde y})],\label{23}\\
    \mathbf{y}&=\mathbf{\tilde{y}}+\mu_1(\overline{\mathbf{y}}-\mathbf{\tilde y}),\label{244}
\end{align}
where $\mu_1{\in}[0, 1]$ can be determined by the Armijo rule, $\nabla f_{\beta}(\mathbf{y})$ denotes the gradient of function $f_{\beta}(\mathbf{y})$, $\Omega$ denotes the constraint set of problem \eqref{p2}, and $P_\Omega[\cdot]$ denotes the projection of the point $(\overline{\mathbf{y}}-\mathbf{\tilde y})$  onto $\Omega$, i.e., the optimal solution to the following equivalent problem,
\begin{subequations}\label{cxh}
\begin{align}
& \mathop{\min}_{\overline {\mathbf{y}}}\parallel \overline {\mathbf{y}}-(\mathbf{\tilde y}-\nabla f_{\beta}(\mathbf{\tilde y})) \parallel^2 \\
&\textrm{s.t.} \quad \eqref{19e}-\eqref{19g},\eqref{22}.
\end{align}
\end{subequations}

Next we show how problem \eqref{cxh} can be globally solved using an efficient bisection method. We note that problem \eqref{cxh} is convex and can be solved by considering its dual problem\cite{cvx_book}. In this regard, we define the partial Lagrangian associated with problem \eqref{cxh} as
\begin{align}
\mathfrak{L}(\overline {\mathbf{y}},\lambda)=\parallel \overline {\mathbf{y}}-(\mathbf{\tilde y}-\nabla f_{\beta}(\mathbf{\tilde y})) \parallel^2+\lambda U(\mathbf{\overline {\mathbf{y}}; \tilde{\mathbf{y}}}),
\end{align}
where $\lambda$ is a Lagrange multiplier. Then, the dual problem \eqref{cxh} can be expressed as
\begin{align}\label{dual2}
    \mathop{\max}_{\lambda\geq 0} h(\lambda)
\end{align}
where $h(\lambda)$ is the dual function given by
\begin{align} \label{dual}
\mathop{\min}_{\overline {\mathbf{y}}} \mathfrak{L}(\overline {\mathbf{y}},\lambda)\quad  \textrm{s.t.} \quad \eqref{19e}-\eqref{19g}.
\end{align}
Note that problem \eqref{dual} can be decomposed into two independent linearly constrained convex quadratic optimization subproblems w.r.t. $\{{P}_{1}^{R},{P}_{2}^{R}\}$ and $\{{P}_{1}^{A},{P}_{2}^{A}\}$, respectively, both of which can be globally solved in closed-form. The detailed derivation can be found in Appendix C.
\begin{algorithm}
\scriptsize
\centering
\caption{Proposed inexact BCD algorithm for problem \eqref{smoothProblem} }\label{table2}
\begin{itemize}
\item [0.] Define the tolerance of accuracy $\zeta$ and the maximum number of iterations $N_\mathrm{max}$. Initialize the algorithm with a feasible point $\mathbf{x}^{0}=[F^{0}_l, F^0_r, \alpha^0,\nu^0,\mathbf{y}^0]^T$. Set the iteration index $i$=0
\item [1.] \textbf{repeat}
\item [2.]\quad Perform bisection method to obtain $F^{i+1}_l$, $F^{i+1}_r$, and $\alpha^{i+1}$, respectively.
\item [3.]\quad Perform line-search method to obtain $\nu^{i+1}$ with Armijo backtracking step \\
\quad size.
\item [5.]\quad Perform one-step PG method to obtain $\mathbf{y}^{i+1}$, using Algorithm 3\\\quad in Appendix C.
\item [6.]\quad  Update the iteration number $i\leftarrow i+1$
\item [7.] \textbf{until} $|  f_{\beta}(\mathbf{x}^{i})- f_{\beta}(\mathbf{x}^{i-1})|\leq \zeta$ or reaching the maximum number of iterations.
\end{itemize}
\end{algorithm}

Based on the above derivation, we summarize the proposed low complexity  IBCD algorithm as Algorithm 2, where the five blocks of variables are sequentially updated. We note that the implementation of the one-step PG method in step 5 involves the use of Algorithm 3 presented in Appendix C. We can show that every limit point, denoted as $\mathbf{x}^{\ast}$, generated by Algorithm 2 is a stationary point of the smoothed problem, i.e., minimizing $f_{\beta}(\mathbf{x})$ subject to \eqref{19b}-\eqref{19h}. The proof is similar to that of \emph{Lemma 1} in \cite{convergence_pg}, and is therefore omitted.  Meanwhile, the computational complexity of Algorithm 3, which dominates the one-step PG method, can be assessed by the number of FPOs $n_2=I_2 I_3 M_2$, where $I_2$ denotes the number of iterations required by the main IBCD loops in Algorithm 2, $I_3$ denotes the number of iterations required by the Algorithm 3{\color{blue}}, and $M_2$ is the number of required FPOs at each iteration of Algorithm 3. Obviously, the value of $M_2$ is far less than $M_1$ in the CCCP based algorithm. Besides, it has been shown in \cite{pg_rate} that the convergence rate of the PG method is $\mathcal{O}(1/I_{2})$.

\section{ Special cases }\label{Special case}
In this section, we investigate problem $\mathbf{P1}$ by considering the special cases $\{\alpha=1,\nu=1\}$ and $\{\alpha=0,\nu=0\}$, corresponding to the DF and AF only relay schemes, respectively. For each one of these special cases, we propose more efficient algorithmic solutions.
\subsection{DF Relay Scheme (case $\alpha = 1$, $\nu=1$)}
Here, we focus on the scenario where user $A$ has very limited computational resource and therefore offloads all its computational tasks to the MERS. The MERS then decodes the computing tasks, executes them using its computational resources, re-encodes the computation results (using possibly a different codebook), and finally transmits the results to user $B$. Hence, user $A$ shares computational results with user $B$, employing only the DF relaying scheme. Substituting $\alpha=1$ and $\nu=1$ into problem $\mathbf{P1}$, we obtain
\begin{subequations}\label{pd9}
\begin{align}
    \mathop{\min}_{F_r,P_{2}^{A},P_{2}^{R}}&~ E_{D}+\gamma t_{D}\label{pd1}\\
    \textrm{s.t.} \quad & 0 < F_r\leq F_{r}^{\mathrm{max}},\label{pd2}\\
      & 0 < P_{2}^{A} \leq P^{\mathrm{max}}_{A}, \label{pd3}\\
      & 0 < P_{2}^{R} \leq P^{\mathrm{max}}_{R},\label{pd4}
\end{align}
\end{subequations}
where $
E_{D}\triangleq L K_r \eta_r {F_r}^2+{L}P_{2}^{A}/{R_{\mathrm{DF1}}}+{\rho L}P_{2}^{R}/{R_{\mathrm{DF2}}},
$
and
$
t_{D}\triangleq {K_r L}/{F_{r}}+{L}/{R_{\mathrm{DF1}}}+{\rho L}/{R_{\mathrm{DF2}}}.
$
By observing that the objective function and the constraints are separable w.r.t. the three blocks of variables, $F_r$, $P_{2}^{A}$, and $P_{2}^{R}$, problem \eqref{pd9} can be decomposed into three independent problems, whose individual solutions are developed below.
\subsubsection{The subproblem w.r.t. $F_{r}$} The variable $F_r$ is updated by solving the following linearly constrained convex problem:
\begin{align}
    &\mathop{\min}_{0 < F_r\leq  F_{r}^{\mathrm{max}}}~ L K_r \eta_r {F_r}^2+\gamma {K_r L}/{F_{r}}
\end{align}
 Applying the first-order optimality condition yields a closed-form solution as follows,
\begin{align}\label{dd1}
    F_r=\min(\mathop{\max} (0,({{\gamma}/{2\eta_r}})^{\frac{3}{2}}), F_{r}^{\mathrm{max}}).
\end{align}
In this case, we note that the optimal computation speed of the MERS, $F_r$, depends on the weighting factor $\gamma$ and the CPU power coefficient $\eta_r$, but is independent of the size $L$ of the computational task.
\subsubsection{The subproblem w.r.t. $P_{2}^{A}$} We update variable $P_{2}^{A}$ by solving the following optimization problem:
\begin{align}\label{pd5}
    &\mathop{\min}_{0 < P_{2}^{A}\leq P^{\mathrm{max}}_{A}}~  {(P_2^A +\gamma)L}/\left(W\log_2\left(1+{ P_{2}^{A}| h_{A}^{(2)}|^2}/{\sigma_{R2}^2}\right)\right)
\end{align}
It can be easily verified that this problem is non-convex, so that its direct solution remains difficult. However, by introducing auxiliary variable $u=1/R_{\mathrm{DF}1}$, with $R_{\mathrm{DF}1} \equiv R_{\mathrm{DF}1}(P_2^A)$ given by (7), problem \eqref{pd5} can be transformed into the following convex optimization problem,
\begin{align}\label{pd6}
    &\mathop{\min}_{u\geq G_2}~ G_1 u(2^{\frac{1}{Wu}}-1)  +\gamma L u
\end{align}
where $G_1\triangleq\frac{L\sigma_{R2}^2}{| h_{A}^{(2)}|^2}$ and $G_2 \triangleq R_{\mathrm{DF}1}(P^{\mathrm{max}}_{A})$. Similar to the proof of \emph{{Lemma 4}} in \cite{7}, we can show that the second-order derivative of the objective function of problem \eqref{pd6} is always greater than or equal to zero. Due to the analytic and convex nature of its objective function,  problem \eqref{pd6} can be efficiently solved by using bisection method \cite{cvx_book}. Given the optimal solution $u^{\star}$ in \eqref{pd6}, the optimal $P_{2}^{A}$ of \eqref{pd5} can be expressed as
\begin{align}\label{dd2}
    P_{2}^{A}={\sigma_{R2}^2}(2^{\frac{1}{Wu^{\star}}}-1)/{| h_{A}^{(2)}|^2}.
\end{align}
\subsubsection{The subproblem w.r.t. $P_{2}^{R}$} The variable $P_{2}^{R}$ is updated by solving the following optimization problem:
\begin{align}\label{pd7}
    &\mathop{\min}_{0 < P_{2}^{R}\leq P^{\mathrm{max}}_{R}}~ {\rho(P+\gamma)L}/\left(W\log_2\left(1+{ P_{2}^{R}| h_{B}^{(2)}|^2}/{\sigma_{B2}^2}\right)\right)
\end{align}
It is seen that problems \eqref{pd7} and \eqref{pd5} have a similar structure and hence, by introducing auxiliary variables $v=1/R_{\mathrm{DF}2}$ and $R_{\mathrm{DF}2} \equiv R_{\mathrm{DF}2}(P_2^R)$ given in (10), \eqref{pd7} can be transformed into a convex optimization problem as follows,
\begin{align}\label{pd8}
    &\mathop{\min}_{v\geq G_4}~ G_3 v(2^{\frac{1}{Wv}}-1)  +\gamma \rho L v
\end{align}
where $G_3\triangleq {\rho L\sigma_{B2}^2}/{| h_{B}^{(2)}|^2}$ and $G_4\triangleq R_{\mathrm{DF}2}(P^{\mathrm{max}}_{R})$. Problem \eqref{pd8} can be globally solved using an efficient bisection method. Given its optimal solution $v^{\star}$, the optimum solution of problem \eqref{pd7} is obtained as
\begin{align}\;\label{dd3}
    P_{2}^{R}={\sigma_{B2}^2}(2^{\frac{1}{Wv^{\star}}}-1)/{| h_{B}^{(2)}|^2}.
\end{align}

\subsection{AF Relay Scheme (case $\alpha=0,\nu=0$):}
Here, we focus on the scenario where the MERS does not provide computing resources and all the computational tasks of user $A$ are performed locally. The MERS then only amplifies the signal received from user $A$ (i.e., the results of the local computations) and forwards it to user $B$. Hence, only the AF relaying scheme is employed in transfer of computational results from $A$ to $B$. Substituting $\alpha=0$ and $\nu=0$ into problem $\mathbf{P1}$, we obtain
\begin{subequations}\label{pa9}
\begin{align}
     \mathop{\min}_{F_l,P_{1}^{A},P_{1}^{R}}&~ E_{A}+\gamma t_{A}\label{pa1}\\
    \textrm{s.t.} \quad & 0 < F_l\leq F_{l}^{\mathrm{max}},\label{pa2}\\
      & 0 < P_{1}^{A} \leq P^{\mathrm{max}}_{A}, \label{pa3}\\
      & P_{1}^{R}\!\sigma_{R1}^2\!+\!| h_{A}^{(1)}\!|^2P_{1}^{R}P_{1}^{A}+P_{2}^{R} \leq P^{\mathrm{max}}_{R},\label{pa4}
\end{align}
\end{subequations}
where$
    E_{A}\triangleq L K_l \eta_l {F_l}^2+(P_{1}^{A}+P_{1}^{R}P_{1}^{A}|h_{A}^{(1)}|^2+P_{1}^{R}\sigma_{R1}^2){\rho L}/{R_{\mathrm{AF}}}
$, and $
t_{A}\triangleq{K_lL}/{F_l}+{\rho L}/{R_{\mathrm{AF}}}.
$
Observing that the constraints are separable w.r.t. the variables, i.e, $F_l$ and $\mathbf{z}\triangleq [P_{1}^{A},P_{1}^{R}]^T$, problem \eqref{pa9} can be decomposed into two independent subproblems, whose respective solutions are derived below.
\subsubsection{The subproblem w.r.t. $F_{l}$} The variable $F_l$ is updated by solving a linearly constrained convex optimization problem as follows,
\begin{align}
    &\mathop{\min}_{0 < F_l\leq F_{l}^{\mathrm{max}}}~ L K_l \eta_l {F_l}^2+\gamma {K_l L}/{F_{l}}
\end{align}
By applying the first-order optimality condition, the following closed-form solution is obtained
\begin{align}\label{aa1}
    F_l=\min(\mathop{\max} (0,({{\gamma}/{2\eta_l}})^{\frac{3}{2}}),F_{l}^{\mathrm{max}}).
\end{align}
\subsubsection{The subproblem w.r.t. $\mathbf{z}$} Let us consider the subproblem w.r.t. $\mathbf{z}$, which is given by
\begin{subequations}\label{pa5}
\begin{align}
&\mathop{\min}_{\mathbf{z}}~f(\mathbf{ z}) \nonumber\\
    &~\textrm{s.t.} \quad  0 < P_{1}^{A} \leq P^{\mathrm{max}}_{A}, \\
      & ~~~~~~~P_{1}^{R}\!\sigma_{R1}^2\!+\!| h_{A}^{(1)}\!|^2P_{1}^{R}P_{1}^{A}+P_{2}^{R} \leq P_{R}^{\mathrm{max}},\label{pa6}
\end{align}
\end{subequations}
where
\begin{align}
f(\mathbf{ z})=&\frac{2\rho L(P_{1}^{A}+P_{1}^{R}P_{1}^{A}|h_{A}^{(1)}|^2+P_{1}^{R}\sigma_{R1}^2)}{W \log_2(1+\frac { P_{1}^{A}P_{1}^{R}|h_{B}^{(1)} h_{A}^{(1)}|^2}{P_{1}^{R}|h_{B}^{(1)}|^2\sigma_{R1}^2+\sigma_{B1}^2})}\nonumber\\
&+\frac{2\gamma\rho L}{W \log_2(1+\frac { P_{1}^{A}P_{1}^{R}|h_{B}^{(1)} h_{A}^{(1)}|^2}{P_{1}^{R}|h_{B}^{(1)}|^2\sigma_{R1}^2+\sigma_{B1}^2})}.\nonumber
\end{align}
It can be observed that problems \eqref{pa5} and \eqref{p2}  have a similar structure. Hence, following the same approach as used for updating the variable block $\bm{y}$ in Section IV, we first approximate \eqref{pa6} as the convex constraint \eqref{22} and apply the \emph{one-step} PG method to \eqref{pa5}. We update $\mathbf{z}$ according to
\begin{align}
    \overline {\mathbf{z}}&=P_\Omega[\mathbf{\tilde z}-\nabla f(\mathbf{\tilde z})],\label{33}\\
    \mathbf{z}&=\mathbf{\tilde{z}}+\mu_2(\overline{\mathbf{z}}-\mathbf{\tilde z}),\label{344}
\end{align}
where $\mu_2{\in}[0, 1]$ can be determined by Armijo rule, $\nabla f(\mathbf{\tilde z})$ denotes the gradient of $f$, $\Omega$ denotes the constraint set of problem \eqref{pa5}, and $P_\Omega[\cdot]$ denotes the projection of the point $(\overline {\mathbf{z}}-\mathbf{\tilde z})$  onto $\Omega$, namely the optimal solution to the following problem
\begin{align}\label{last}
&\mathop{\min}_{\mathbf z}\parallel \mathbf{z}-(\mathbf{\tilde z}-\nabla f(\mathbf{\tilde z})) \parallel^2 \nonumber\\
&\textrm{s.t.} \quad \eqref{22}, \, \eqref{pa5}.
\end{align}
The remaining details of the derivation are omitted due to space considerations.
\section{Simulation Results}\label{simulation_section}
In this section, we use Monte Carlo simulations to demonstrate the benefits of the proposed CCCP-based and low complexity IBCD algorithms for RACO systems in terms of the end-to-end delay and system energy consumption. The simulations are run on a desktop computer with (Intel i7-920) CPU running at
$2.66$ GHz and $24$ Gbytes RAM, while the simulation parameters are set as follows unless specified otherwise. All the channel gains are independently generate based on a Rayleigh fading model with average gain factor $\sigma_h^2=E[|h|^2]=10^{-3}$. The radio bandwidth available for data transmission from user $A$ to user $B$ via the MERS is $W=40$ MHz for the combination of the AF and DF schemes. The background noise at MERS and user B is $-169$ dBm/Hz.
The maximum transmit power levels of user $A$ and the MERS are set to $P^{\mathrm{max}}_{A}=1$ Watts and $P^{\mathrm{max}}_{R}=5$ Watts, respectively. The maximum computation speed of user $A$ and the MERS are $F^{\mathrm{max}}_{l}=200$ MHz and $F^{\mathrm{max}}_{r}=600$ MHz, respectively. For user $A$, the data size of the tasks before computation follows a uniform distribution over the interval $[1\cdot10^5, 5\cdot10^5]$ bits, the conversion ratio is fixed to $\rho=0.1$, and the required number of CPU cycles per bit for both user $A$ and the MERS is set to $K\triangleq K_l=K_r=10^3$ cycles/bit. Furthermore, the power consumption coefficients for the given chip architecture are set as $\eta \triangleq \eta_l=\eta_r=10^{-28}$ (Watts$ \times \text{s}^3$)\cite{7,vtc_relay,wcnc_maxmin}.  In the implementation of the IBCD algorithm, the smoothness factor $\beta$ in (53) is set to $10$. All results are obtained by averaging over 100 independent channel realizations. For convenience, the simulation parameters are listed in Table \ref{Simulation_parameter}.
\begin{table}[t]
 \scriptsize
  \centering
  \caption{Simulation parameters}\label{Simulation_parameter}

 \begin{tabular}{|l|r|}
   \hline
   \textbf{Parameters} & \textbf{Value} \\
   \hline
   Radio bandwidth of AF and DF subchannels $W$ & $40$ MHz\\
   \hline
   Average gain factor $\sigma_h^2$ & $10^{-3}$\\
     \hline
   Maximum transmission power of user $A$ $P_{A}^{\text{max}}$&$1$ Watts\\
       \hline
       Maximum transmission power of the MERS $P_{R}^{\text{max}}$&$5$ Watts\\
        \hline
       Maximum computation capacity of user $A$ $F^{\mathrm{max}}_{l}$&$200$ MHz\\
       \hline
       Maximum computation capacity of the MERS $F^{\mathrm{max}}_{r}$&$600$ MHz\\
       \hline
       Size of tasks before computation $L$&$[1\times10\cdot5, 5\cdot10^5]$ bits\\
       \hline
       CPU cycles required to process a bit $K$& $1000$ cycles/bit\\
       \hline
       Conversion ratio $\rho$ & 0.1\\
       \hline
       Coefficient depending on chip architecture $\eta$ & $10^{-28}$ Watts$\times \text{s}^3$,\\
       \hline
       Smoothness factor $\beta$& 10\\
       \hline
 \end{tabular}
 \end{table}

\subsection{Convergence Performance}
\begin{figure}[!t]
\centering \scalebox{0.32}{\includegraphics {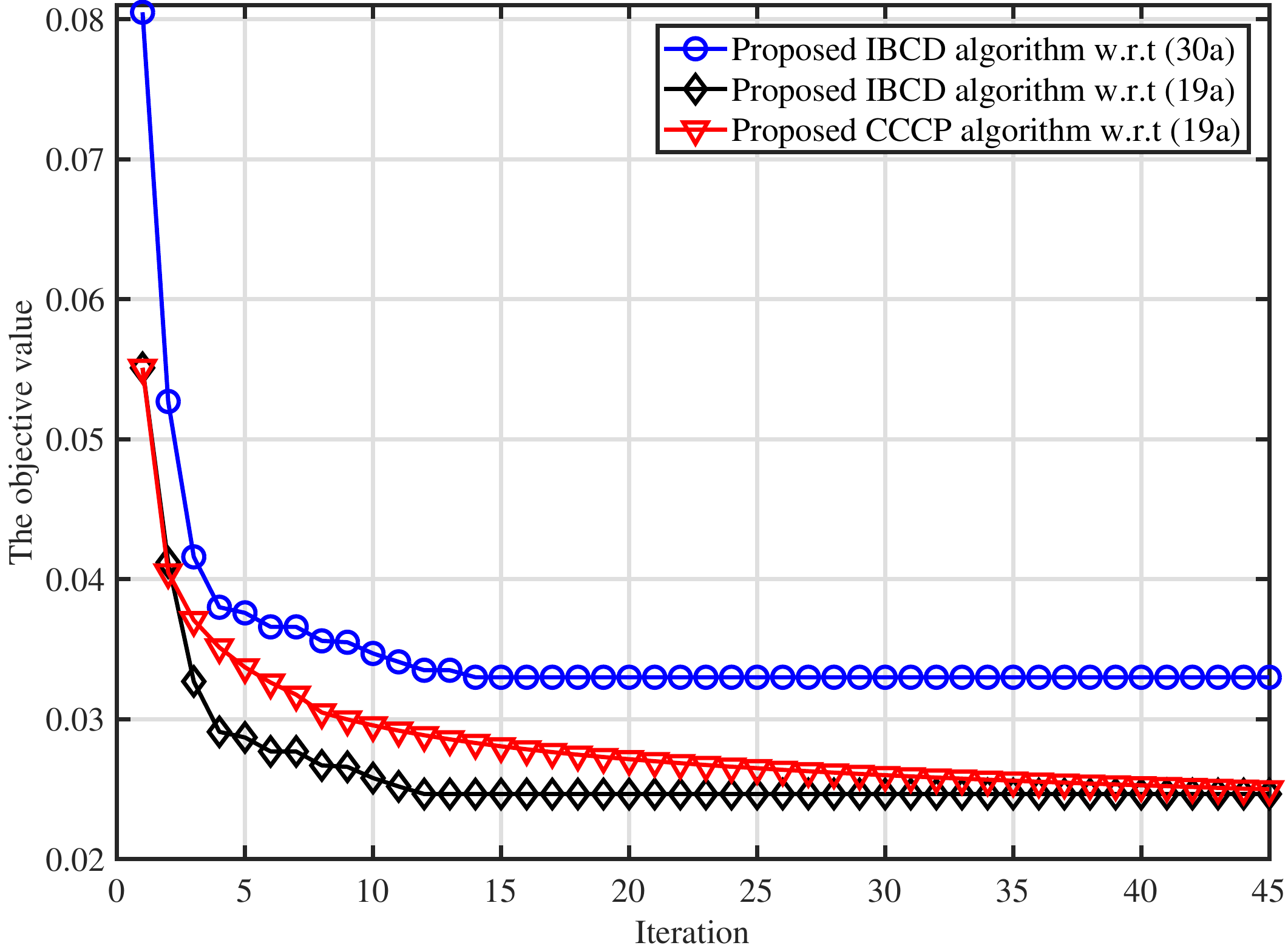}}
\caption{ Convergence behavior for the proposed algorithms for the case $\gamma=0.01$ (J$\cdot \text{sec}^{-1}$).}\label{fig:f1}
\end{figure}
We begin by studying the convergence performance of the proposed CCCP-based and low-complexity IBCD algorithms. For the CCCP-based algorithm, Fig. \ref{fig:f1} plots the values of the objective function \eqref{WSDEM} versus the iteration number, while for the IBCD algorithm, the values of both the objective function \eqref{WSDEM} and its smoothed approximation \eqref{smoothProblem} versus iteration number are plotted. These curves reveal that despite the existence of a gap between the two objective functions (i.e., green versus red lines), the low-complexity IBCD algorithm monotonically converges to the same value as that achieved by the CCCP-based algorithm, (i.e., green versus blue lines). We note that the IBCD algorithm can achieve faster convergence than the CCCP-based algorithm. In addition, since the CCCP-based algorithm requires solving a sequence of complex convex problems,  a single iteration of IBCD runs much faster than the corresponding CCCP iteration, as shown by the average run time data in Table \ref{CPUtime}. Hence, the IBCD algorithm is more efficient than the CCCP-based algorithm.
 \begin{table}[!t]
 \scriptsize
  \centering
  \caption{Comparison of execution times for CCCP-based and IBCD algorithms.}\label{CPUtime}
 \begin{tabular}{|c|c|c|}
   \hline
   & Average number of  & Average execution\\
   & iterations to converge & time per iteration\\
   \hline
   IBCD &20&$0.1$ s\\
   \hline
   CCCP &$45$&$1.82$ s\\
   \hline
 \end{tabular}
 \end{table}
\subsection{Performance of Proposed Algorithms for General Case: HR Scheme}
\begin{figure}[!ht]
\centering
	\subfloat[]{\centering \scalebox{0.33}{\includegraphics{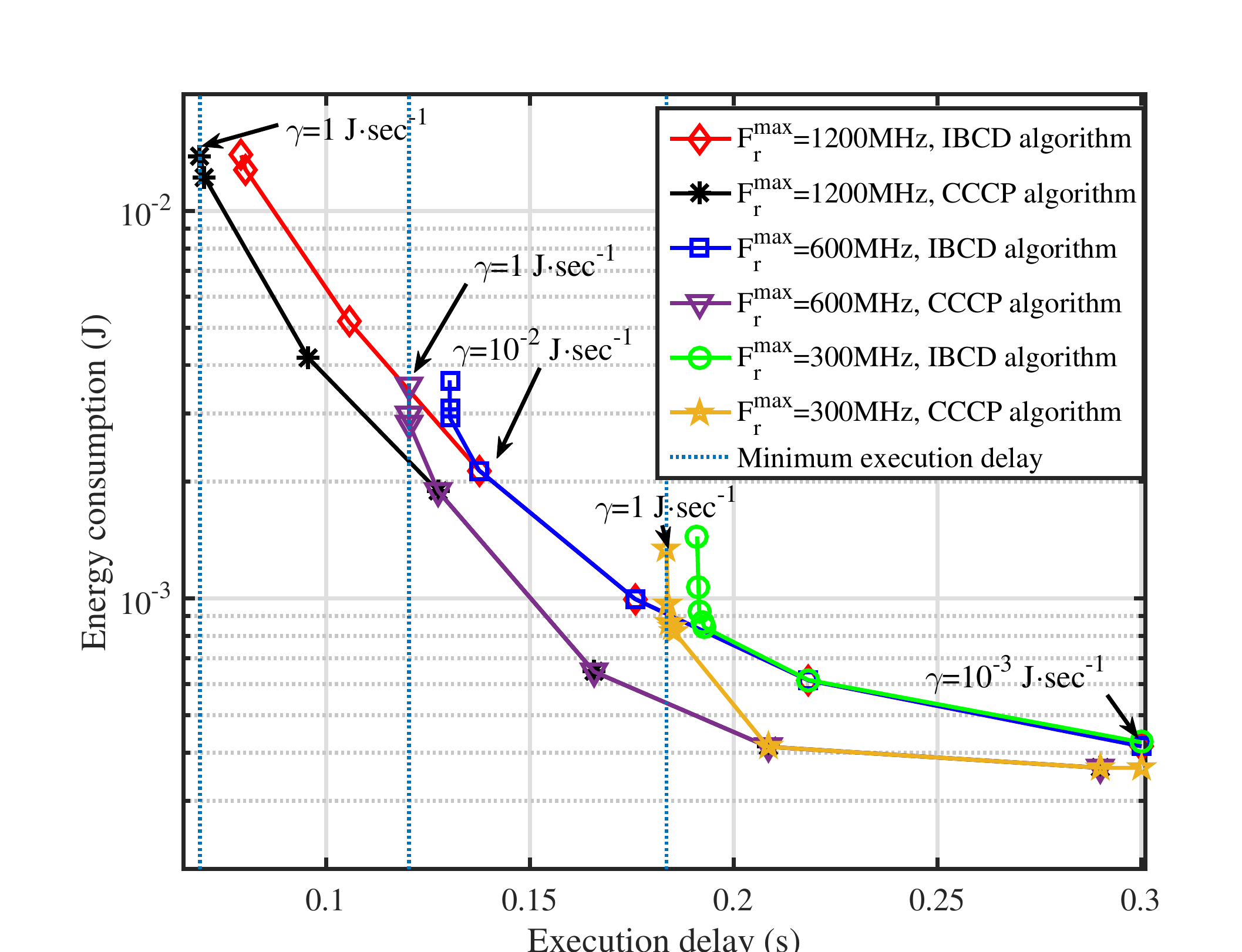}} \label{fig3:power}}\\
	\subfloat[]{\centering \scalebox{0.35}{\includegraphics{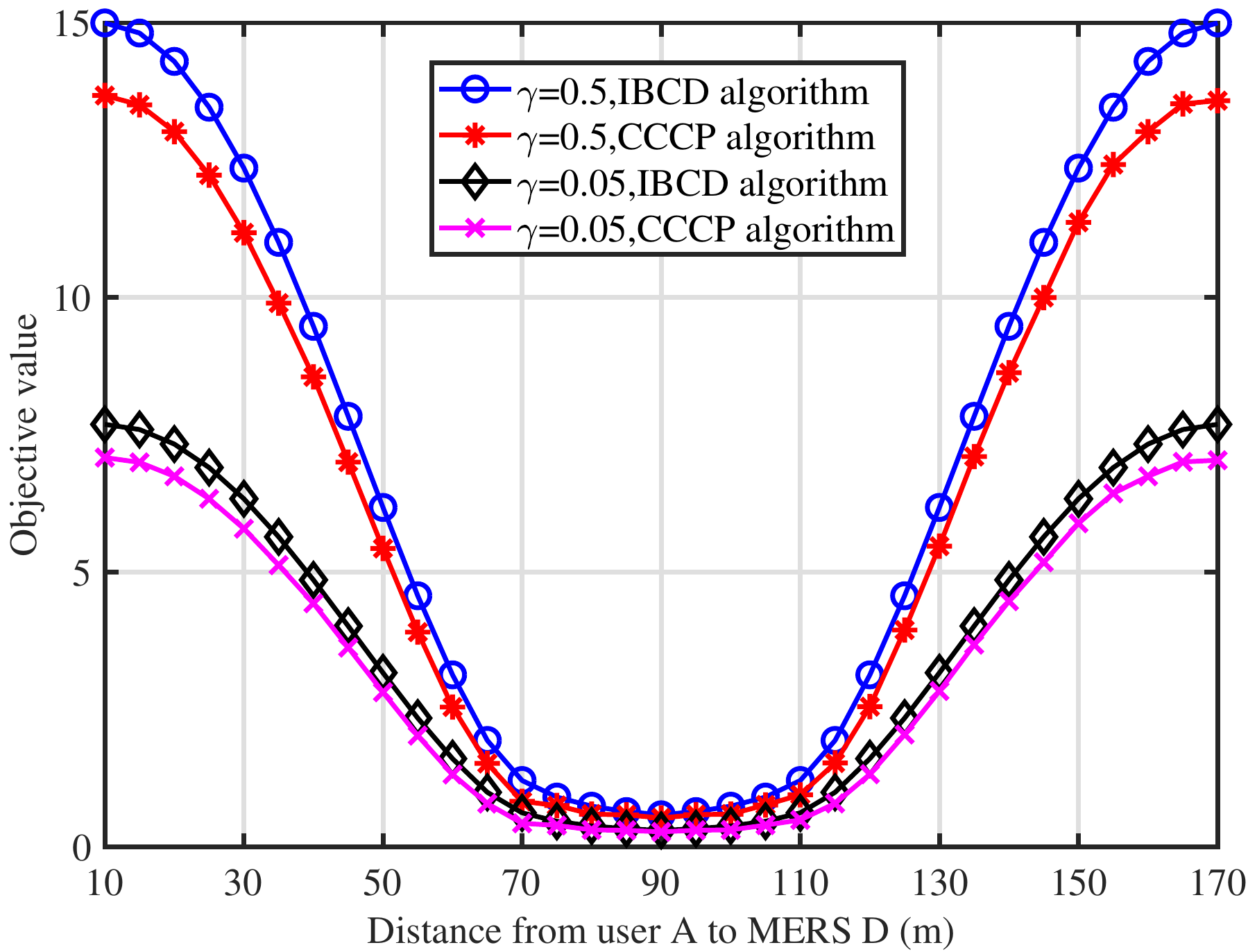}} \label{fig3:finite}}	
	\caption{(a) Energy consumption versus execution delay for different algorithms; (b) objective function value versus the distance between user $A$ and MERS for different algorithms. }
	\label{Fig3}
	\end{figure}

We now investigates the performance of the proposed resource allocation algorithms when applied to the general case of RACO system with HR architecture. The tradeoff between the system energy consumption and execution delay for the CCCP-based and IBCD algorithms is illustrated in Fig. \ref{fig3:power} for $F_{r}^{\max}=300$, $600$, and $1200$ MHz. It can be observed that the energy consumption increases while the execution delay decreases as the weighting parameter $\gamma$ in (19) increases. When $\gamma$ is relatively large, our approach gives more weight to the delay minimization; consequently, our proposed algorithms can achieve the minimum execution delay (i.e., vertical blue dashed line). Conversely, when $\gamma$ is relatively small, our approach gives more emphasis to the minimization of the energy consumption minimization and, in particular, tends to yield the same energy consumption irrespective of the value of $F_{r}^{\max}$. This is because the minimum energy consumption is achieved when $F_r$ is very small [cf. (16)]. This reveals a fundamental design principle for RACO systems: when our design emphasis is on the minimization of  energy consumption, there is no need to deploy too excessive resources at the MERS. Besides, we also note from Fig. \ref{fig3:power} that the performance of the IBCD algorithm is inferior to that of the CCCP-based algorithm, especially when $\gamma$ is relatively large. In effect, it appears that the smooth approximation of execution delay $t_{\mathrm{sys}}$ in (35) leads to a notable performance loss when more weight is given to delay minimization. Even so, the IBCD algorithm  is still very promising due to its lower computational complexity as demonstrated earlier.
%

Fig. \ref{fig3:finite} shows the values of the objective function (19) versus the distance between user $A$ and the MERS, with $\gamma=0.5$ and $0.05$. In this simulation, we consider a practical scenario where the distance between user $A$ and user $B$ is set to 180 meters(m), with the MERS being located on the line segment joining them at a distance $D$ from user $A$, where $10\leq D \leq170$m. The following path loss model is employed for the wireless links: $\mathrm{PL}_0 (\frac{d}{d_0})^{-\chi}$, where $d$ denotes the distance between the transmitter and receiver, $\chi=3$ is the  path loss exponent, and $\mathrm{PL}_0=-60$ dB denotes the path loss at a reference distance of $d_0=10$m \cite{110}. It can be observed that the objective function first decreases to reach a minimum at $D=90$m and then increases as $D$ further increases. This is due to the fact that with increasing $D$, the wireless channel between user $A$ and the MERS becomes weaker, while that between the MERS and  user $B$ becomes stronger. This benefits the forwarding of computational results from the MERS to user $B$, but incurs increased energy consumption and transmission delay from user $A$ to the MERS. These observations are consistent with the theoretical analysis in \cite{110}, where the optimal location for the relay is shown to be at the middle point between user $A$ and $B$.


\subsection{Performance of Proposed Algorithms for Special Cases: AF and DF Schemes Only}
These special cases occur in the limit $\{\alpha=0, \nu=0\}$ and $\{\alpha=1,\nu=1\}$: in the former case, user $A$ performs all of its computational tasks locally and sends the results to user $B$ using the AF relaying scheme; while in the latter case, user $A$ offloads all of its tasks to the MERS for mobile edge execution using the DF scheme. In these two cases, problem $\mathbf{P1}$ reduces to the simpler forms \eqref{pa9} and \eqref{pd9} respectively, which in turn lead to the more efficient algorithmic solutions developed in Section \ref{Special case}, which we now investigate. In addition to the proposed HR architecture, the following baseline schemes are considered for comparison:
\begin{itemize}
\item [1)] The proposed HR architecture with time division (TDHR): The locally computed results and the offloaded raw tasks are each transmitted using the complete available radio spectrum but different time slots. The durations of time slots for AF and DF scheme are the same. In addition, the locally computed results are transmitted just after the offloaded raw task.
\item [2)] The proposed HR architecture with frequency division (FDHR): This scheme employs both AF and DF relaying over two orthogonal frequency bands. Moreover, the AF and DF scheme occupy the same channel bandwidth of $W/2=20$ MHz.
\end{itemize}
%
%
\begin{figure}[!t]
\centering \scalebox{0.32}{\includegraphics {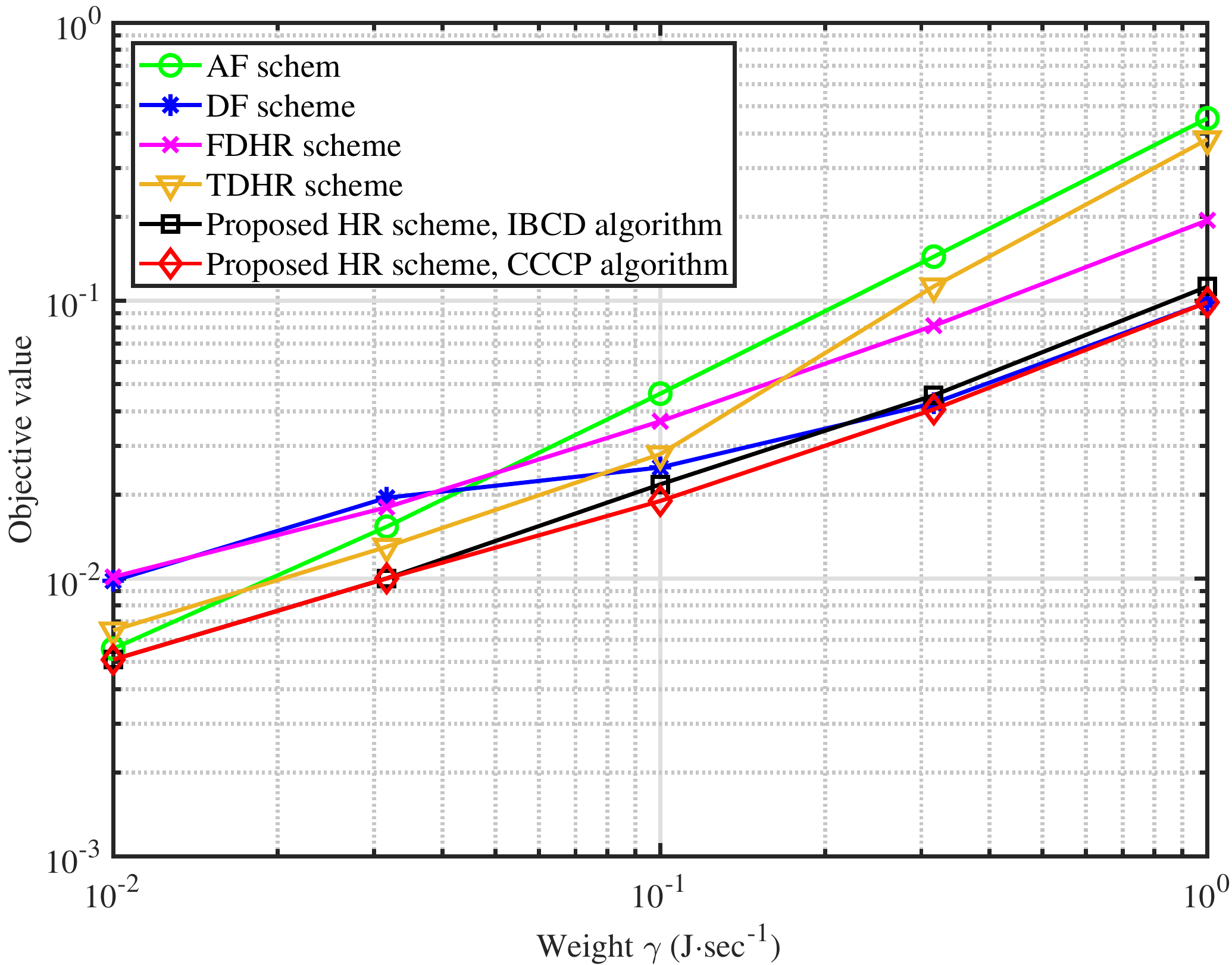}}
\caption{ Objective function value versus the weight factor between execution delay and energy consumption for different relay schemes.}\label{fig:f4}
\end{figure}

Fig. \ref{fig:f4} illustrates the objective function (19) versus the weight factor $\gamma$ between execution delay and energy consumption for different schemes and algorithms. It can be observed that for all schemes under comparison, the value of the objective function increases with $\gamma$. First, we discuss the performance comparison between the proposed HR scheme and the AF and DF only schemes. It is seen that the DF only scheme outperforms the AF scheme for larger $\gamma$. This indicates that when more emphasis is given to the minimization of the delay, the performance of the RACO system can be further improved by employing the DF scheme, as compared to the AF scheme. When $\gamma$ is relatively small and  more weight is given to the energy consumption minimization, the AF scheme outperforms its DF counterpart. Besides, it is interesting to note that the HR scheme with CCCP-based optimization can achieve smaller weighted sum of execution delay and energy consumption performance than both the AF and DF schemes for all possible values of $\gamma$. These results demonstrate the effectiveness of the HR architecture in handling different scenarios for user  preferences (i.e. weighting factor $\gamma$) and its ability to strike a better balance between energy minimization and execution delay, thereby endowing added flexibility to the RACO system. Moreover, it is observed that the performance of the AF scheme is close to that of the proposed HR scheme at relatively smaller $\gamma$, while the performance of the DF scheme is enhanced  monotonically and coincides with the proposed HR scheme when $\gamma= 1$. Second, we compare the performance of the HR schemes with different division modes. It should be emphasized that the TDHR scheme tends to be superior at smaller $\gamma$ but performs worse at larger $\gamma$. This is because the TDHR scheme allocates two different time slots for transmitting the locally computed results and the offloaded raw tasks, which is inefficient when more emphasis is given to the minimization of the delay. As for the FDHR scheme, its performance is also inferior due to its fixed and ineffective spectrum utilization method. Finally, it is observed that the proposed HR architecture achieves significant gain over the FDHR and the TDHR schemes for all $\gamma$ regime, because both the frequency resources and time resources are fully utilized by the proposed HR architecture.


%
%

\begin{figure}[!t]
\centering
	\subfloat[]{\centering \scalebox{0.32}{\includegraphics{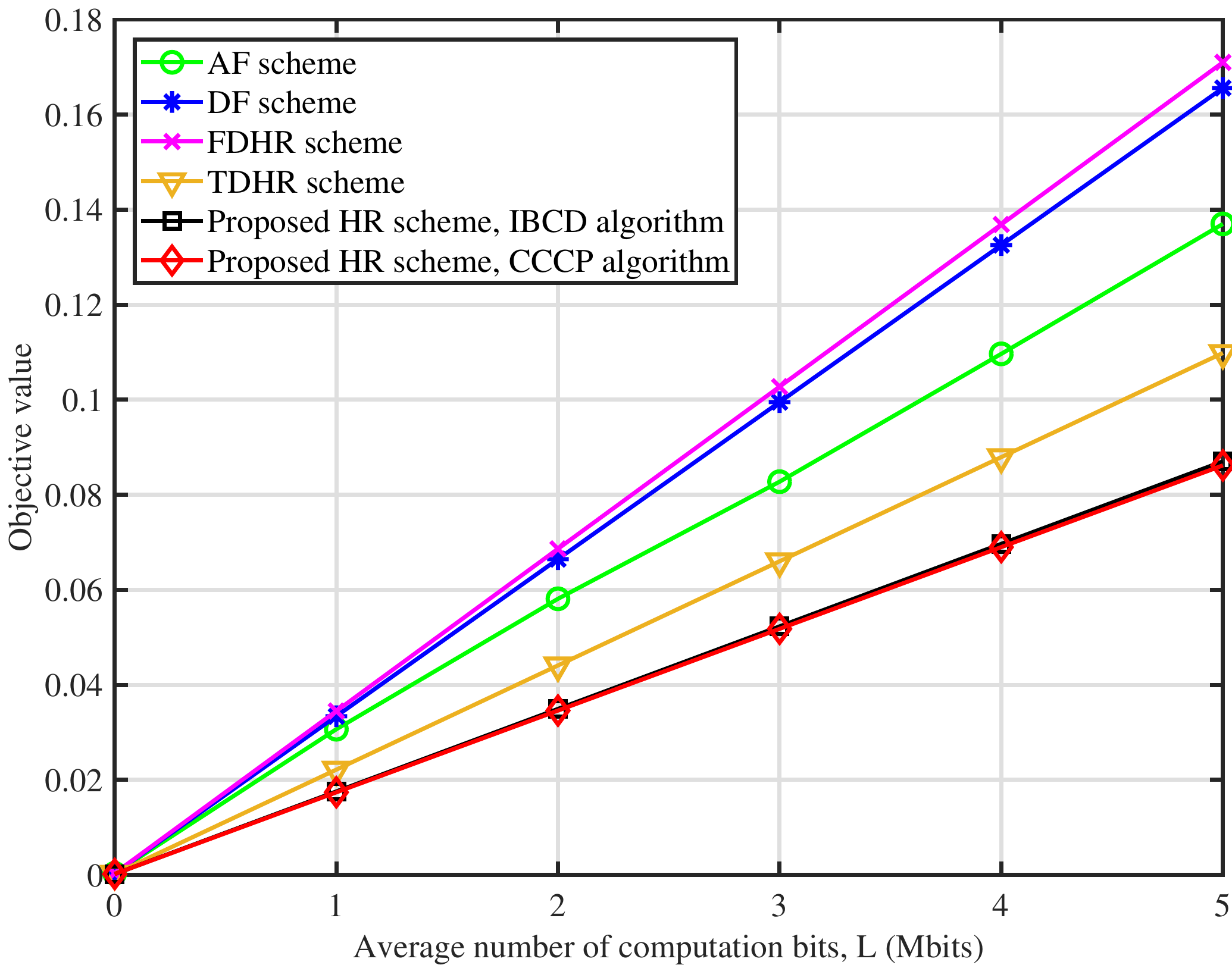}} \label{fig5:power}}\\
	\subfloat[]{\centering \scalebox{0.32}{\includegraphics{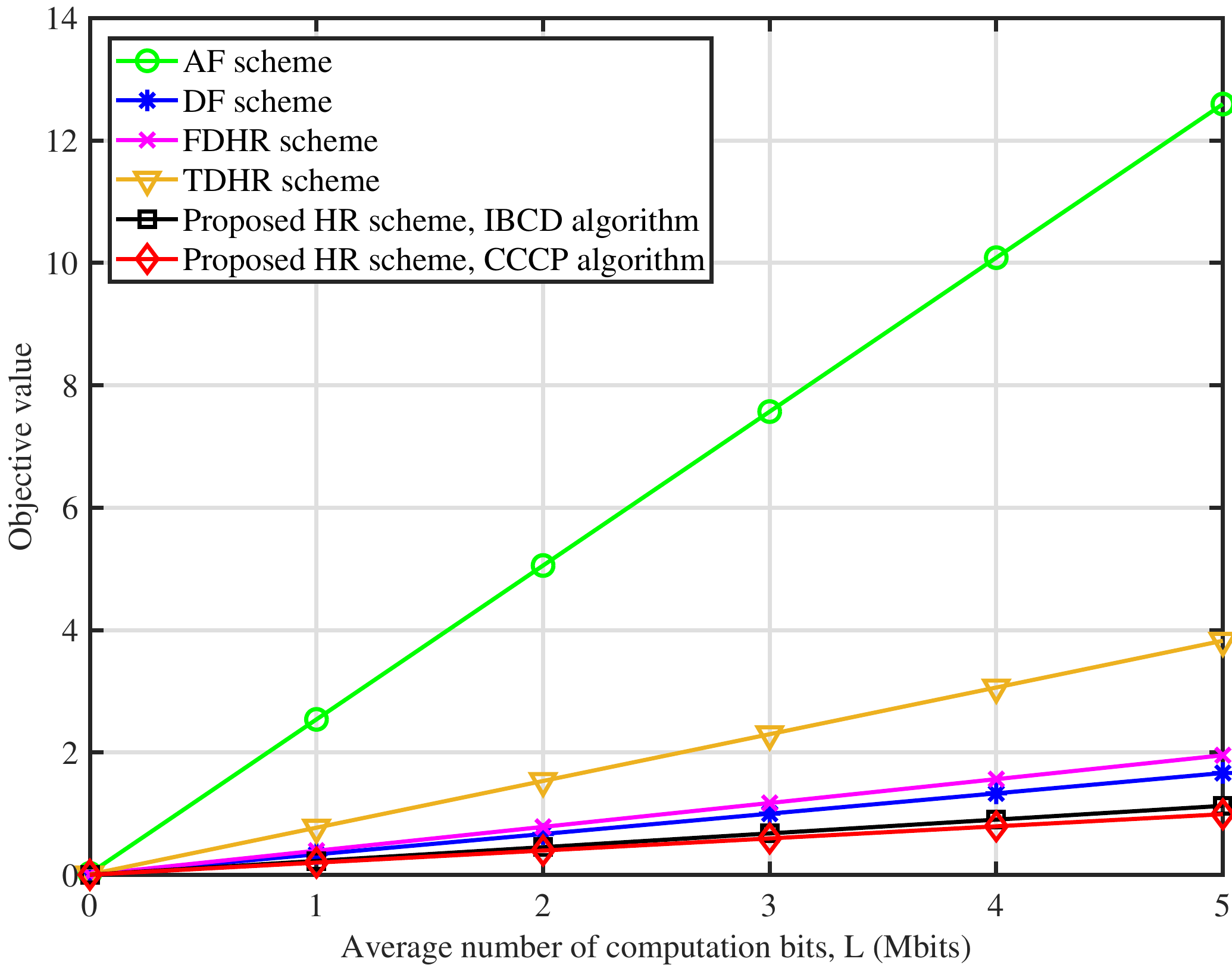}} \label{fig5:finite}}	
	\caption{Objective function value versus  (a) the average number of computation bits for $\gamma=0.01$ (J$\cdot \text{sec}^{-1}$); (b) the average number of computation bits for $\gamma=1$ (J$\cdot \text{sec}^{-1}$). }
	\label{Fig5}
\end{figure}

Fig. \ref{fig5:power} and Fig. \ref{fig5:finite} show plots of the objective function (19) versus the average number of computation bits $L$ for different schemes/algorithms and for two different choice of weighting factor, i.e., $\gamma=0.01$ and $\gamma=1$, respectively. It can be observed that the objective function values of all schemes and for the two different choices of $\gamma$ linearly with respect to the number of computation bits $L$. This is intuitively satisfying since more computation bits pose more stringent requirements on both computation and radio resources, which incurs increased energy consumption and time delay for the RACO system. Note that in this paper, we aim to minimize the weighted sum of execution delay and energy consumption by appropriately allocating both the computational and communication resources. For such a scenario, the tradeoff between execution delay and energy consumption relies more on the inner computational characteristics of tasks, and is invariant to the input size $L$. Consequently, the optimum value of the offloading ratio is independent of $L$. In Fig. \ref{fig5:power}, the performance of the AF scheme is superior to that of the DF one, but still not as good as that of the HR architecture.  This implies that the AF scheme is more suitable for implementation than the DF scheme when the design emphasis is placed on the minimization of energy consumption. In Fig. \ref{fig5:finite}, the performance of the DF scheme is much better than the AF one, and coincides with that of the HR architecture. Hence, the DF scheme is more suitable than the AF scheme when the design emphasis is placed on the minimization of the delay. According to  \eqref{aa1}, the optimal computational speed of user $A$ depends on the weight factor and the CPU's coefficient, but not on the size of the computational tasks. Hence, it can be inferred that an increase of the weight factor $\gamma$ will pose more stringent requirements on the computation speed of user $A$. When the optimal computational speed exceeds the allowed maximum for user $A$, the execution delay of the RACO system will be severely penalized, thereby leading to to a poor performance. This further demonstrates the advantages of cooperative computation offloading between user $A$ and the MERS. It is also interesting to note from Fig. \ref{Fig5} that the HR architecture with CCCP-based optimization outperforms both the TDHR scheme and the FDHR scheme in terms of the weighted sum of execution delay and energy consumption performance, which demonstrates the importance of the bandwidth allocation and relay strategy.

\setcounter{equation}{61}
\begin{figure*}[b]
\hrulefill
\begin{align}
f_2(\mathbf{x}, \bm{\phi})\triangleq &\frac{1}{2}[L K_r \eta_r (\alpha^2+{F_r}^4)+L K_l \eta_l  (\alpha+{F_l}^2)^2+({P_{1}^{A}})^2+{t_{\mathrm{A}}}^2+\sigma_{R1}^2(({P_{1}^{R}})^2+{t^2_{\mathrm{A}}})\nonumber\\
&+| h_{A}^{(1)}|^2(s_1^2+{t^2_{\mathrm{A}}})+({P_{2}^{A}})^2+{t_{\mathrm{D1}}}^2+({P_{2}^{R}})^2+{t^2_{\mathrm{D2}}}].\label{appB_1}
\end{align}
\end{figure*}

\setcounter{equation}{62}
\begin{figure*}[b]
\hrulefill
\begin{align}
\hat f_2(\mathbf{x}, \bm{\phi};\tilde{\mathbf{x}},\tilde{\bm{\phi}})=&\frac{1}{2}[L K_r \eta_r(\tilde\alpha^2+\tilde{F_r}^4)+L K_l \eta_l(\tilde\alpha+\tilde{F_l}^2)^2+\!L K_r \eta_r \tilde\alpha(\!\alpha\!-\!\tilde\alpha)+2L K_r \eta_r \tilde{F_r}^3(\!F_r\!-\!\tilde{F_r})+L K_l \eta_l(\tilde\alpha+\tilde{F_l}^2)(\alpha-\tilde\alpha)\nonumber\\
&+2L K_l \eta_l\tilde{F_l}(\tilde\alpha+\tilde{F_l}^2)(F_l-\tilde{F_l})]+(\tilde{P_{1}^{A}})^2+{\tilde t_{\mathrm{A}}}^2+\sigma_{R1}^2((\tilde{P_{1}^{R}})^2+{\tilde t_{\mathrm{A}}}^2)+| h_{A}^{(1)}|^2(\tilde s_1^2+{\tilde t_{\mathrm{A}}}^2)\!+\!({\tilde P_{2}^{A}})^2\!+\!{\tilde t_{\mathrm{D1}}}^2\!+\!({\tilde P_{2}^{R}})^2\nonumber\\
&+{\tilde t_{\mathrm{D2}}}^2]+[\tilde P_{1}^{A}(P_{1}^{A}-\tilde P_{1}^{A})+\tilde t_{\mathrm{A}}(t_{\mathrm{A}}-\tilde t_{\mathrm{A}})+\sigma_{R1}^2(\tilde P_{1}^{R}(P_{1}^{R}-\tilde P_{1}^{R})+\tilde t_{\mathrm{A}}(t_{\mathrm{A}}-\tilde t_{\mathrm{A}}))+| h_{A}^{(1)}|^2(\tilde s_1(s_1-\tilde s_1)\nonumber\\
&+\tilde t_{\mathrm{A}}(t_{\mathrm{A}}-\tilde t_{\mathrm{A}}))+\tilde P_{2}^{A}(P_{2}^{A}-\tilde P_{2}^{A})+\tilde t_{\mathrm{D1}}(t_{\mathrm{D1}}-\tilde t_{\mathrm{D1}})+\tilde P_{2}^{R}(P_{2}^{R}-\tilde P_{2}^{R})+\tilde t_{\mathrm{D2}}(t_{\mathrm{D2}}-\tilde t_{\mathrm{D2}})].\label{appB_2}
\end{align}
\end{figure*}

\setcounter{equation}{65}
\begin{figure*}[b]
\hrulefill
\begin{align}
D_1\triangleq\left({1+\lambda| h_{A}^{(2)}|^2}\right)^{-1}({ \tilde{P}_{1}^{R}-{\partial f_{\beta}(\mathbf{\tilde y})}/{\partial P_1^{R}}}-\lambda\sigma_{R1}^2+\lambda| h_{A}^{(1)}|^2\tilde{P}_{1}^{R}),\quad D_2\triangleq{-\left({1+\lambda| h_{A}^{(2)}|^2}\right)^{-1}\lambda| h_{A}^{(1)}|^2}.\label{lemma_D1}
\end{align}
\end{figure*}

\setcounter{equation}{67}
\begin{figure*}[b]
\hrulefill
\begin{align}
A_1&\triangleq{1+D_2^2+\lambda | h_{A}^{(1)}|^2(1+D_2)^2},\quad C_2\triangleq\tilde{P}_{2}^{A}-{\partial f_{\beta}(\mathbf{\tilde y})}/{\partial P_2^{A}}, \nonumber\\
C_1&\triangleq\frac{\tilde{P}_{1}^{A}-\frac{\partial f_{\beta}(\mathbf{\tilde y})}{\partial P_1^{A}}+D_2(\tilde{P}_{1}^{R}-\frac{\partial f_{\beta}(\mathbf{\tilde y})}{\partial P_1^{R}}-D_1-\lambda \sigma^2_{R1})+\lambda |h_{A}^{(1)}|^2(D_2\tilde{P}_{1}^{R}+\tilde{P}_{1}^{A}-(1+D_2)D_1)}{1+D_2^2+\lambda| h_{A}^{(1)}|^2(\!1\!+\!D_2)^2}.\label{lemma_AC}
\end{align}
\end{figure*}
\setcounter{equation}{56}
\section{Conclusion}\label{conclusion_section}
This paper has investigated the problem of joint cooperative relaying and computation sharing within a MEC context, where the aim is to minimize the weighted sum of the execution delay and the energy consumption in MEC systems. To support RACO, we proposed a HR architecture which combines the merits of AF and DF relaying to enhance performance. To tackle the challenging optimization problem under consideration, where the design variables are highly coupled through the objective and constraint functions, an efficient CCCP-based algorithm was proposed to jointly optimize the bandwidth allocation, the transmit power level of the source user $A$ and the MERS, the computational resources and the percentage of computational tasks offloaded through the DF relay channel, under constraints on the CPU speed and the transmission power budgets at user $A$ and M-RES. To address the difficulty arising from the high computational complexity of the CCCP-based algorithm, we  proposed a simplified algorithm based on a smoothed approximation along with the IBCD method, that takes advantage of the problem structure to reduce complexity. We then considered two special situations corresponding to limiting cases of the offloading ratio, i.e., AF and DF scheme, and for these proposed efficient solutions. Numerical results showed that the proposed HR architecture can achieve better performance than the AF and the DF scheme, and is particularly well-suited for RACO applications due to its great flexibility along with reduced execution delay and energy consumption. Due to the space limitation, there have been various important issues that have not been addressed in this paper, e.g. the
more general case with multiple user, bi-directional communication, and so on.


\appendices
\section{Derivation Of Equivalent Transformation}\label{appendix_A}
First, let us focus on the term $E_{\mathrm{sys}}$ in problem $\mathbf{P1}$. By introducing auxiliary variables $t_{\mathrm{A}},t_{\mathrm{D1}}$ and $t_{\mathrm{D2}}$, corresponding to the upper bounds of the transmitting time delay in each relay scheme, we can move the associated mathematical expressions to constraints. Similarly, we introduce the auxiliary variables $s_1$ corresponding to the upper bound of $P_{1}^{A}P_{1}^{R}$ in \eqref{19i}. The resultant equivalent optimization problem is then given by
\begin{subequations}\label{ut0}
\begin{align}
     \mathop{\min}_{\mathbf{x}, \bm{\phi}}&~  E_{\mathrm{sys}}+\gamma   t_{sys}\\
    \textrm{s.t.} \quad &\eqref{19b}-\eqref{19h}, \\
    &\frac{2(1-\alpha)\rho L}{(1-\nu)W\log_2\left(1+\frac { P_{1}^{A}P_{1}^{R}|h_{B}^{(1)} h_{A}^{(1)}|^2}{P_{1}^{R}|h_{B}^{(1)}|^2\sigma_{R1}^2+\sigma_{B1}^2}\right)} \leq t_{\mathrm{AF}}\label{ut1}\\
    &\frac{\alpha L}{\nu W\log_2\left(1+\frac{ P_{2}^{A}| h_{A}^{(2)}|^2}{\sigma_{R2}^2}\right)}\leq t_{\mathrm{D1}}\label{ut2}\\
    &\frac{\alpha \rho L}{\nu W\log_2\left(1+\frac{ P_{2}^{R}| h_{B}^{(2)}|^2}{\sigma_{B2}^2}\right)} \leq t_{\mathrm{D2}}\label{ut3}\\
    & P_{1}^{A}P_{1}^{R} \leq s_1,
    P_{1}^{R}\sigma_{R1}^2+ |h_{A}^{(1)}|^2 s_1+P_{2}^{R} \leq P^{\mathrm{max}}_{R}.\label{ut7}
\end{align}
\end{subequations}
In order to transform \eqref{ut1}-\eqref{ut3} into a simpler form, we introduce an additional set of auxiliary variables, i.e., $\{R_{\mathrm{A}},R_{\mathrm{D1}},R_{\mathrm{D2}}, \Lambda_{\mathrm{A}},\Lambda_{\mathrm{D1}}, \Lambda_{\mathrm{D2}},
\varphi_1,\varphi_2,\varphi_3,s_2\}$.  Here, we focus on the required manipulations for \eqref{ut1} as an example:
\begin{align}
    R_{\mathrm{A}}\leq {(1-\nu)W}\Lambda_{\mathrm{A}}/{2}\leq {(1-\nu)W}\log_2(1+ {1}/{\varphi_1})/{2},\nonumber\\
    \frac {1}{\varphi_1}\leq \frac { s_2|h_{B}^{(1)} h_{A}^{(1)}|^2}{P_{1}^{R}|h_{B}^{(1)}|^2\sigma_{R1}^2+\sigma_{B1}^2}\leq \frac { P_{1}^{A}P_{1}^{R} |h_{B}^{(1)} h_{A}^{(1)}|^2}{P_{1}^{R}|h_{B}^{(1)}|^2\sigma_{R1}^2+\sigma_{B1}^2}.\nonumber
\end{align}
It should be emphasized that the auxiliary variable $s_2$ is introduced as the lower bound of $P_{1}^{A}P_{1}^{R}$. Then, \eqref{ut0} can be formulated as the equivalent problem below:
\begin{subequations}\label{ut6}
\begin{align}
     \mathop{\min}_{\mathbf{x}, \bm{\phi}}&~  E_{\mathrm{sys}}+\gamma   t_{\mathrm{sys}}\\
    \textrm{s.t.} \quad
    &
    (1-\alpha)\rho L\leq t_{\mathrm{A}}R_{\mathrm{A}},
    2 R_{\mathrm{A}}\leq (1-\nu)W \Lambda_{\mathrm{A}} ,\\
    & \Lambda_{\mathrm{A}} \leq \log_2(1+{1}/{\varphi_1}),
    \alpha L\leq t_{\mathrm{D1}}R_{\mathrm{D1}},\\
    & R_{\mathrm{D1}}\leq \nu W \Lambda_{\mathrm{D1}},
     ~\Lambda_{\mathrm{D1}} \leq \log_2(1+{1}/{\varphi_2}),\\
     &\alpha \rho L\leq{R_{\mathrm{D2}}}t_{\mathrm{D2}},
     ~R_{\mathrm{D2}}\leq \nu W \Lambda_{\mathrm{D2}},\\
     &\Lambda_{\mathrm{D2}} \leq \log_2(1+{1}/{\varphi_3}),
     \sigma_{R2}^2-\varphi_2 P_{2}^{A} |h_{A}^{(2)}|^2\leq 0,\\
     & |h_{B}^{(1)}|^2\sigma_{R1}^2P_{1}^{R}+\sigma_{B1}^2-|h_{B}^{(1)} h_{A}^{(1)}|^2 \varphi_1 s_2\leq 0,\\
     &|h_{B}^{(1)}|^2\sigma_{R1}^2P_{1}^{R}+\sigma_{B1}^2-|h_{B}^{(1)} h_{A}^{(1)}|^2 \varphi_1 s_2\leq 0,\\
    & \sigma_{B2}^2-\varphi_3 P_{2}^{R} |h_{B}^{(2)}|^2\leq 0, s_2 \leq P_{1}^{A}P_{1}^{R}\leq s_1,\\
     &\eqref{19b}-\eqref{19h}.
\end{align}
\end{subequations}
Finally, we must address the difficulty posed by the term $t_{\text{sys}}$ in \eqref{ut0}. To this end, we can move the term $t_{\text{sys}}$ into the constraints and introduce the auxiliary variables $\{t_s, t_l,t_r\}$, now corresponding to the upper bound of the whole execution time, local execution time, and edge execution time respectively, which yields equivalent yet more tractable constraints as follows,
\begin{align}
& t_{l}+t_{\mathrm{A}} \leq t_{s}, \, t_{\mathrm{D1}}+t_{r}+t_{\mathrm{D2}}\leq t_{s}\label{ut8}\\
& t_{l}\geq{K_l(1-\alpha)L}/{F_{l}}, \, t_{r}\geq {K_r\alpha L}/{F_{r}}.\label{ut9}
\end{align}
Substituting \eqref{ut8} and \eqref{ut9} in problem \eqref{ut6}, we obtain the equivalent problem $\mathbf{P2}$. Complete this proof.

\section{Derivation Of \eqref{f30}}\label{appendix_B}
We note that $f_3(\mathbf{x}, \bm{\phi})$ is nonconvex due to the products of optimization variables. To tackle this nonconvexity by applying the CCCP, we first need to transform $f_3(\mathbf{x}, \bm{\phi})$ into a difference-of-convex (DC) program. Focusing on the operation of the last term $P_{2}^{R}t_{\mathrm{D2}}$ in $f_3(\mathbf{x}, \bm{\phi})$ as an example, we have:
\begin{align}
        P_{2}^{R}t_{\mathrm{DF2}}=\frac{1}{2} [ (P_{2}^{R}+t_{\mathrm{D2}})^2-({P_{2}^{R}})^2-{t^2_{\mathrm{D2}}}]
\end{align}
We can follow the same approach to handle the remaining nonconvex terms in $f_3(\mathbf{x}, \bm{\phi})$, and finally obtain $
 f_3(\mathbf{x}, \bm{\phi})\triangleq f_1(\mathbf{x}, \bm{\phi})-f_2(\mathbf{x}, \bm{\phi}),
$
where $f_1(\mathbf{x}, \bm{\phi})$ and $f_2(\mathbf{x}, \bm{\phi})$ are respectively defined in \eqref{top1_1} and \eqref{appB_1}.

Based on the CCCP concept \cite{cccp}, we approximate the convex function $f_2(\mathbf{x}, \bm{\phi})$ in the $i$th iteration by its first order Taylor expansion around the current point $(\tilde{\mathbf{x}},\tilde{\bm{\phi}})$ as \eqref{appB_2}.

Therefore, using the above results we can obtain a locally tight upper bound for the objective function of problem $\mathbf{P2}$, i.e.,
$
    f_3(\mathbf{x}, \bm{\phi};\tilde{\mathbf{x}},\tilde{\bm{\phi}})\triangleq f_1(\mathbf x,\bm{\phi})-\hat f_2(\mathbf{x}, \bm{\phi};\tilde{\mathbf{x}},\tilde{\bm{\phi}}).
$

 \section{Solving problem \eqref{dual} for updating $\bm{\overline{y}}$ }\label{appendix_C}
In this part, we derive each step of the update procedure for solving \eqref{dual}.
 \subsubsection{Subproblem w.r.t $\{{P}_{1}^{R},{P}_{2}^{R}\}$} By applying the first-order optimality condition, we obtain a closed-form solution as follows,
 \setcounter{equation}{63}
\begin{align}
&\overline{P}_1^{R}=\min(\mathop{\max} (0,D_1+D_2P_1^{A}),P_R^{\mathrm{max}}),\label{pp1}\\
&\overline{P}_2^{R}=\min (\mathop{\max} (0,\tilde{P}_{2}^{R}-{\partial f_{\beta}(\mathbf{\tilde y})}/{\partial P_2^{R}}-\lambda),P_R^{\mathrm{max}})\label{pp2},
\end{align}
where both $D_1$ and $D_2$ are defined \eqref{lemma_D1} in as displayed at the bottom of the next page.

\subsubsection{Subproblem w.r.t $\{{P}_{1}^{A},{P}_{2}^{A}\}$} Substituting \eqref{pp1} and \eqref{pp2} into \eqref{dual}, we can obtain a quadratic optimization problem w.r.t $\{{P}_{1}^{A},{P}_{2}^{A}\}$ as follows
 \setcounter{equation}{67}
\begin{subequations}\label{PP3}
\begin{align}
 \mathop{\min}_{P_{1}^{A},P_{2}^{A}}&A_1(P_1^{A}-C_1)^2+(P_2^{A}-C_2)^2 \\
\textrm{s.t.} \quad &\eqref{19e}, \, \eqref{19g}
\end{align}
\end{subequations}
where all the term $A_1$, $C_1$, and $C_2$ are defined in \eqref{lemma_AC} in as displayed at the bottom of the next page.
Solving problem \eqref{PP3} is equivalent to computing a projection of the point $(C_1,C_2)$ onto the set $\Omega_1=\{(P_{1}^{A},P_{2}^{A})\mid\ P_{1}^{A}+P_{2}^{A}\leq P^{\mathrm{max}}_{A}, P_{1}^{A}\geq 0, P_{2}^{A}\geq 0\}$. When the point $(C_1,C_2)$ is inside of the feasible region, the optimal solution to problem \eqref{PP3} is immediately obtained as $\overline{P}_1^{A}=C_1,\overline{P}_2^{A}=C_2$. When the point $(C_1,C_2)$ is not inside of the feasible region, the optimal solution to \eqref{PP3} can be obtained on the boundary of the feasible region. In other words, the constraint \eqref{19e} is strictly satisfied, i.e.,
$
P_2^{A}=P_{\mathrm{max}}^{A}-P_1^{A}.
$
By substituting it into \eqref{PP3}, we can obtain a quadratic formulation w.r.t $P_1^{A}$:
 \setcounter{equation}{69}
\begin{align}\label{PP5}
\mathop{\min}_{\{0\leq P_1^{A} \leq  P^{\mathrm{max}}_{A}\}} ~A_1(P_1^{A}-C_1)^2+(P^{\mathrm{max}}_{A}-C_2-P_1^{A})^2
\end{align}
Applying the first-order optimality condition yields a closed-form solution as follows
 \begin{align} \label{t1}
    \overline{P}_1^{A}=\min(\mathop{\max}(0,({A_1+1})^{-1}({A_1C_1+P^{\mathrm{max}}_{A}-C_2})),P^{\mathrm{max}}_{A}).
 \end{align}
Given the optimal solution $\overline{P}_1^{A}$ in \eqref{PP5}, the optimal $\overline{P}_2^{A}$ can be expressed as
\begin{align} \label{t2}
\overline{P}_2^{A}=P^{\mathrm{max}}_{A}-\overline{P}_1^{A}.
\end{align}
Furthermore, since the objective function of problem \eqref{dual}  given $\lambda > 0$ is strictly convex, problem \eqref{dual} has a unique solution. It follows that $h(\lambda)$ is differentiable
for $\lambda > 0$ and its derivative is $U(\mathbf{\overline{y}; \tilde{\mathbf{y}}})$. Consequently, the dual problem \eqref{dual2} can be efficiently solved using the bisection method, which is summarized in Algorithm 3.
\begin{algorithm}
\scriptsize
\centering
\caption{Proposed bisection algorithm for the problem \eqref{cxh}}\label{table3}
\begin{itemize}
\item [0.]Input: the current point $\tilde{P}_{1}^{A},\tilde{P}_{2}^{A},\tilde{P}_{1}^{R}$ and $\tilde{P}_{2}^{R}$.
\item [1.]Define the lower and upper bounds of the Lagrange multiplier respectively $\lambda_u$ and $\lambda_l$, and the tolerance of accuracy $\epsilon$. Initialize the algorithm with $0\leq \lambda_l<\lambda_u$.
\item [2.]\textbf{Repeat}
\item [3.]\quad Let $\lambda\leftarrow \frac{\lambda_l+\lambda_u}{2}$.
\item [4.]\quad  Update $\overline{{P}}_{1}^{R}$ and $\overline{{P}}_{2}^{R}$ according to \eqref{pp1} and \eqref{pp2}, respectively.
\item [5.]\quad\textbf{If} point $(C_1,C_2)$ is inside of the feasible region, \\ \quad \quad \quad update $\overline{{P}}_{1}^{A}=C_1$ and $\overline{{P}_{1}}^{A}=C_2$, respectively;
\item [6.]\quad\textbf{else} update $\overline{{P}}_{1}^{A}$ and $\overline{{P}}_{2}^{A}$ according to \eqref{t1} and \eqref{t2}, respectively.
\item [7.]\quad\textbf{If} $U(\mathbf{\overline{y}; \tilde{\mathbf{y}}})\geq 0$, let $\lambda_l\leftarrow \lambda$; \textbf{else} let $\lambda_u\leftarrow \lambda$.
\item [8.]\textbf{Until} $\mid \lambda_u-\lambda_l \mid \leq \epsilon$.
\end{itemize}
\end{algorithm}

\end{document}